\documentstyle[12pt,epsfig]{article}
\def\aprge{\buildrel > \over {_{\sim}}}

\title {Matter effects  in long--baseline experiments,  \\
 the flavor content  of the heaviest (or lightest) neutrino \\
and the  sign  of $\Delta m^2$}

\author{Paolo Lipari \\
I.N.F.N., sezione di Roma, and \\
Dipartimento di Fisica, Universit\`a di Roma ``la Sapienza",\\
P. A. Moro 2,  I-00185 Roma, Italy}

\begin{document}

\maketitle

\begin{abstract}
The  neutrinos  of  long baseline beams travel  
inside the Earth's  crust where  the  density is 
$\rho \simeq 2.8$~g~cm$^{-3}$.
If electron  neutrinos  participate in the oscillations,
matter  effects will modify  the oscillation  probabilities
with respect to   the vacuum case.
Depending on the sign of $\Delta m^2$ 
an  MSW  resonance will exist 
for  neutrinos  or anti--neutrinos with energy 
$E_\nu^{res} \simeq  4.7 \cdot |\Delta m^2|/(10^{-3}$~eV$^2$)~GeV.
For  $\Delta m^2$ in the interval indicated  by 
the Super--Kamiokande experiment 
this  energy range is  important for the 
proposed long baseline   experiments.

For  positive $\Delta m^2$ the most important effects of  
matter  are a  9\% (25\%)  enhancement 
of the transition probability $P(\nu_\mu \to \nu_e)$
for the KEK to  Kamioka  (Fermilab to  Minos and CERN to
Gran Sasso) beam(s)
in  the  energy region where the  probability has its first maximum,
and  an approximately equal 
suppression of $P(\overline{\nu}_\mu \to \overline{\nu}_e)$.
For  negative $\Delta m^2$ the  effects for neutrinos and
anti--neutrinos   are interchanged.
Producing  beams  of  neutrinos and antineutrinos
and  measuring the   oscillation probabilities  for
both the $\nu_\mu \to \nu_e$ and
$\overline{\nu}_\mu \to \overline{\nu}_e$ transitions
can  solve the sign ambiguity in the determination of $\Delta m^2$.
\end{abstract}

\section {Introduction}
The data   on  atmospheric neutrinos  collected  by Super--Kamiokande
\cite{SK,SK-new}  and other  detectors (Kamiokande, IMB, Soudan and MACRO
\cite{Kamioka,IMB,Soudan,MACRO})
give  good  evidence for the existence of  neutrino oscillations
with $|\Delta m^2| \simeq 10^{-3}$--$10^{-2}$~eV$^2$.
The  experimental   results  show  that
the flux  of  muon  (anti--)neutrinos    is  suppressed 
with respect to the standard--model  prediction
while the flux of  electron (anti--)neutrinos
is  compatible  with the no--oscillation  prediction
taking into account measurement  errors  and  systematic  uncertainties.
This    suggests  that   the dominant  effect of the oscillations
is  $\nu_\mu \to \nu_\tau$ transitions, in good agreement 
with the results of the Chooz  experiment  \cite {Chooz} 
that  put  strong limits  on  possible  $\nu_e \to \nu_x$  
transitions  in the   relevant region of $L/E_\nu$.

It is possible  that the observed disappearance
of the muon (anti--)neutrinos is  due  to oscillation
into a  light sterile  state \cite{sterile},  
and also  other forms   of  `new   physics'
beyond the standard model have  been  proposed
as explanations  of  the atmospheric neutrino data
(see \cite{exotic}    for a  critical  discussion).
In this work  we  will however only consider  
standard  oscillations between three neutrino  flavors.
This   is the simplest  extension of the standard model
that  can describe the data, and  can 
do it very successfully \cite{SK,SK-new}.
In this  framework  one expects 
transitions between  all flavors
($\nu_\mu \leftrightarrow \nu_\tau$, 
$\nu_\mu \leftrightarrow \nu_e$, 
$\nu_e \leftrightarrow \nu_\tau$), and  a  very important  goal  of
future  experiments  will be   the 
measurement  of  (or the setting of more stringent  limits on)
the transitions  involving   electron neutrinos.

Future  experiments   measuring the
disappearance  of  reactor  neutrinos  with longer 
pathlengths (in particular the 
Kamland detector \cite{Kamland}) will   be able to
study   $\nu_e \to \nu_x$  transitions 
down to lower  values of  $\Delta m^2$, however, 
because  of systematic  uncertainties 
\cite {Chooz},  it will be  difficult to  extend  
the  sensitivity of  reactor  experiments 
to values  of the mixing much  lower  that  those already  excluded 
by Chooz.
Higher  statistics    measurement   of the atmospheric  neutrino  fluxes
have a good  potential to search for  the flavor transitions
of the $\nu_e$'s  \cite{Fogli,ADLS,Petcov}  searching for  up--down
asymmetries in the  $e$--like events.  
Also experiments \cite{industry} 
using long  baseline (LBL) neutrino  beams  
such as the KEK  to Kamioka \cite{K2K},
Fermilab to Minos  \cite{Minos} and  CERN to Gran  Sasso
\cite{CERN} projects  have  a  very interesting  potential 
for measuring    $\nu_\mu \to \nu_e$   transitions.
The  LBL beams  are mostly  composed  of $\nu_\mu$'s
with a small $\nu_e$ contamination  below the 1\% level,
and  a detector  with good electron identification  capability,
collecting  a sufficiently  large sample  of events,   can  
be  sensitive to $\nu_\mu \to \nu_e$  transitions
even for  values  of the mixing well below the Chooz  limit.

Because of the sphericity of the  earth, the neutrinos  
of LBL  beams  travel a few kilometers  below the Earth's surface, 
in a medium that  can   be  considered as approximately
homogeneous   with  an electron  density
$n_e \simeq 1.69\cdot 10^{24}$~cm$^{-3}$
(corresponding to a  density $\rho \simeq 2.8$~g~cm$^{-3}$ and 
an electron  fraction $Y_e = n_e/(n_p + n_n) = 0.5$).
If the electron neutrinos   participate in the  oscillations,
the presence of matter  modifies  the oscillation  probabilities.
In this  paper we will  discuss  in some  detail the matter  effects  and
their observable  consequences.

The  work is  organized as  follows:
in the  next  section we  introduce the theoretical framework
used   in this   analysis (a  single relevant $\Delta m^2$);
in section~3  we  discuss  the   existing limits   
on the   three independent parameters 
present in    this framework  ($\Delta m^2$  and  two  mixing parameters);
in section~4  we  compute the neutrino effective  squared  masses and
mixing  matrix  in matter; in section~5  we discuss the  propagation of
neutrinos  in matter  with a  constant electron density
(a  good  approximation  for  LBL beams);
in  section~6  we  finally discuss the  oscillation  probabilities
for  LBL  beams. A   summary is given  in  section~7.

\section {One mass scale  approximation}
In the  general  case of  mixing  between  three  neutrino  flavors, 
one has  to consider three  masses $m_1$, $m_2$ and  $m_3$,
and  a $(3 \times 3)$  unitary mixing   matrix $U$  that  relates the 
flavor
$\{|\nu_e\rangle,  |\nu_\mu\rangle,   |\nu_\tau\rangle \}$
and mass 
$\{|\nu_1\rangle,  |\nu_2\rangle,   |\nu_3\rangle \}$
eigenstates:
\begin {equation}
 |\nu_\alpha \rangle = U_{\alpha j} \; |\nu_j\rangle
\end{equation}
Without  lack  of  generality  we can label 
the mass eigenstates  so that:
\begin {equation}
|m_3^2 - m_2^2| > |m_2^2 - m_1^2|
\end{equation}
and 
\begin{equation}
  m_1^2 \le m_2^2 \le m_3^2 ~~~~{\rm or}  ~~~~~
  m_1^2 \ge m_2^2 \ge m_3^2
\end{equation}
With this  choice $|\nu_3\rangle$  is  the 
`most  isolated' state separated   by the largest  mass
difference  gap  from the neutrino  closest in mass
and is therefore the
heaviest  (or lightest) neutrino, correspondingly
$|\nu_1\rangle$ is the  lightest  (or  heaviest)    neutrino  state.

In this  work   we will  make the approximation 
that a  single  mass  scale is  important for the
experiments we are   considering.
More  explicitely we  will assume  that:
\begin {enumerate}
\item  the squared  mass differences   
$\Delta m^2_{12}$ and $\Delta m^2_{23}$ are  of different  order of magnitude:
\begin {equation}
|m_3^2 - m_2^2| \gg  |m_2^2 - m_1^2|,
\end{equation}

\item the mass  difference $|m_2^2 - m_1^2|$ is  too small
to   give observable  effects    in the  measurements 
considered.
\end{enumerate}
The study of oscillations under these assumptions  has  been 
performed  by several authors in the past 
\cite{Pantaleone,Fogli-onemass,Cardall}.
Formally   it corresponds to  the study of neutrino oscillation
under the  hypothesis  that two neutrinos  are degenerate in mass
($m_2 \simeq m_1$).
The use of the one mass  scale  approximation 
is    motivated by   three type of considerations:
\begin {enumerate}
\item  Neutrino oscillations  in this   approximation
can  still  produce  transitions   between all  flavors,   
and  therefore this is a  much more  general  framework  
than the special  case of two--flavor   mixing. 
The  description of oscillations 
remains    however  much simpler  than in  the  general case,  and therefore 
the one mass scale  approximation
represents  a natural `minimum model' to study oscillations.

\item  In  most theoretical  models 
it is  natural  to expect that the   squared mass differences
are organized in a  hierarchical   form 
$|m_3^2 - m_2^2| \gg |m_2^2 - m_1^2|$.

\item Finally and  most  important, there   are strong  experimental
indications, coming from experiments on solar   neutrinos
\cite{solar} that the mass   difference $\Delta m^2_{12}$ 
controls the observed  suppression of the   solar  neutrino  fluxes, and
is of order $\sim 10^{-5}$~eV$^2$   (for  the MSW   solutions)
or $\sim 10^{-11}$~eV$^2$   (for  the just--so solution).
These low values of  $\Delta m^2_{12}$
are unobservable in long--baseline  experiments
and  can result  in at most  
small  effects  \cite {Peres-Smirnov,Giunti} for low  energy
atmospheric   neutrinos.
Therefore if  oscillations  are the  solution for 
both the atmospheric  and the solar neutrino problems 
the one mass  scale approximation is a  completely adequate framework
to  describe    long--baseline  experiments,
and  is  a  reasonable model for the study of   atmospheric  neutrinos.
\end {enumerate}

When two neutrinos  are  degenerate in mass 
the mixing  matrix  $U$ is  not completely defined.
In fact   the $j$-th column    of  the  matrix
$(U_{e j}, U_{\mu j}, U_{\tau j})$  
describes  the flavor  content  of  the  mass eigenstate  
$|\nu_j\rangle$, and    the  two 
states with the  same mass 
$|\nu_1\rangle$  and  $|\nu_2\rangle$  can
be  chosen in an  infinite  number  of ways.
The  degeneracy  is broken 
when  neutrinos  propagate in 
matter,  because the  $\nu_e$  and  $\nu_\mu$ ($\nu_\tau$)  have
different    potentials (see next section), and in the 
subspace  of  the neutrino  states orthogonal  to  $|\nu_3\rangle$
the  states with  and  without  a $|\nu_e\rangle$  component
have different    effective masses.
It is therefore  convenient to 
choose    one  of the eigenstates,
for  example the $|\nu_1\rangle$ state,  as  orthogonal  to
the flavor  state $|\nu_e\rangle$,  this  also  determines
$|\nu_2\rangle$  (modulo some  phase convention)
as the  normalized  state orthogonal to  $|\nu_1\rangle$  and $|\nu_3\rangle$.
With  these  conventions the  mixing  matrix  $U$  can be   parametrized as:
\begin {equation}
U = \left [
 \begin{array} {c c c}
  0 &  \cos \theta  & \sin \theta  \\
 \cos \varphi & -\sin \theta  \sin \varphi 
               & \cos \theta \sin \varphi \\
 -\sin \varphi  & -\sin \theta \cos \varphi 
               & \cos \theta \cos \varphi 
 \end {array}
\right]
\label{eq:mix0}
\end{equation}
with the two  mixing  angles $\theta$  and  $\varphi$  
defined in the  interval $[0, \pi/2]$.
The  definition of the  mixing matrix  in equation  (\ref{eq:mix0}) 
contains  some arbitrary sign  conventions
(only the absolute  values   $|U_{\alpha j}|$ of  the elements are
uniquely  defined),  the  important point is  that it is    possible
to  define all  elements as  real and  all $CP$ or $T$ violating effects 
vanish  in  the one mass scale approximation.
In the  following we will  also  use
the notation
\begin{equation}
p_{\alpha 3} = |\langle \nu_\alpha | \nu_3\rangle|^2 = |U_{\alpha 3}|^2
\end{equation}
where $p_{\alpha 3}$  is the probability
that the neutrino $|\nu_3\rangle$ has the flavor 
$|\nu_\alpha\rangle$.
These  probabilities obviously satisfy the relation:
\begin {equation}
p_{e 3} + p_{\mu 3} + p_{\tau 3} = 1
\end{equation}

In  the  one mass scale approximation   all  
oscillation effects   (also  in the presence of matter) 
can  be expressed as  a  function of  $\Delta m^2 =  m_3^2 - m_1^2$ 
and of the three probabilities $p_{e 3}$, $p_{\mu 3}$ and $p_{\tau 3}$.
The transition probabilities  in  vacuum  can be written as:
\begin{equation}
P^{vac}_{\nu_\alpha \to \nu_\beta} (L/E_\nu) 
= 4 \;p_{\alpha 3} \;p_{\beta 3}
~\sin^2 \left [ {\Delta m^2 L \over 4\, E_\nu } \right ]
\label{eq:vac-trans}
\end {equation}
and the survival  probabilities as:
\begin{equation}
P^{vac}_{\nu_\alpha \to \nu_\alpha} (L/E_\nu) 
= 1 - 4\;p_{\alpha 3} \;(1-p_{\alpha 3}) 
~\sin^2 \left [ {\Delta m^2 L \over 4\, E_\nu } \right ],
\label{eq:vac-surv}
\end {equation}
For   fixed   $E_\nu$ 
all   probabilities   oscillate  with a  single 
oscillation length $\lambda_0 (E_\nu)$:
\begin {equation}
\lambda_0 (E_\nu)  = { 4 \pi \,E_\nu \over |\Delta m^2| } \simeq 2.470 ~
{E_\nu({\rm GeV}) \over \Delta m^2 ({\rm eV}^2) }~{\rm km}
\label {eq:lambda0}
\end {equation}
that  grows  linearly with increasing $E_\nu$.

\section {Experimental results}

\begin{figure} [t]
\centerline{\psfig{figure=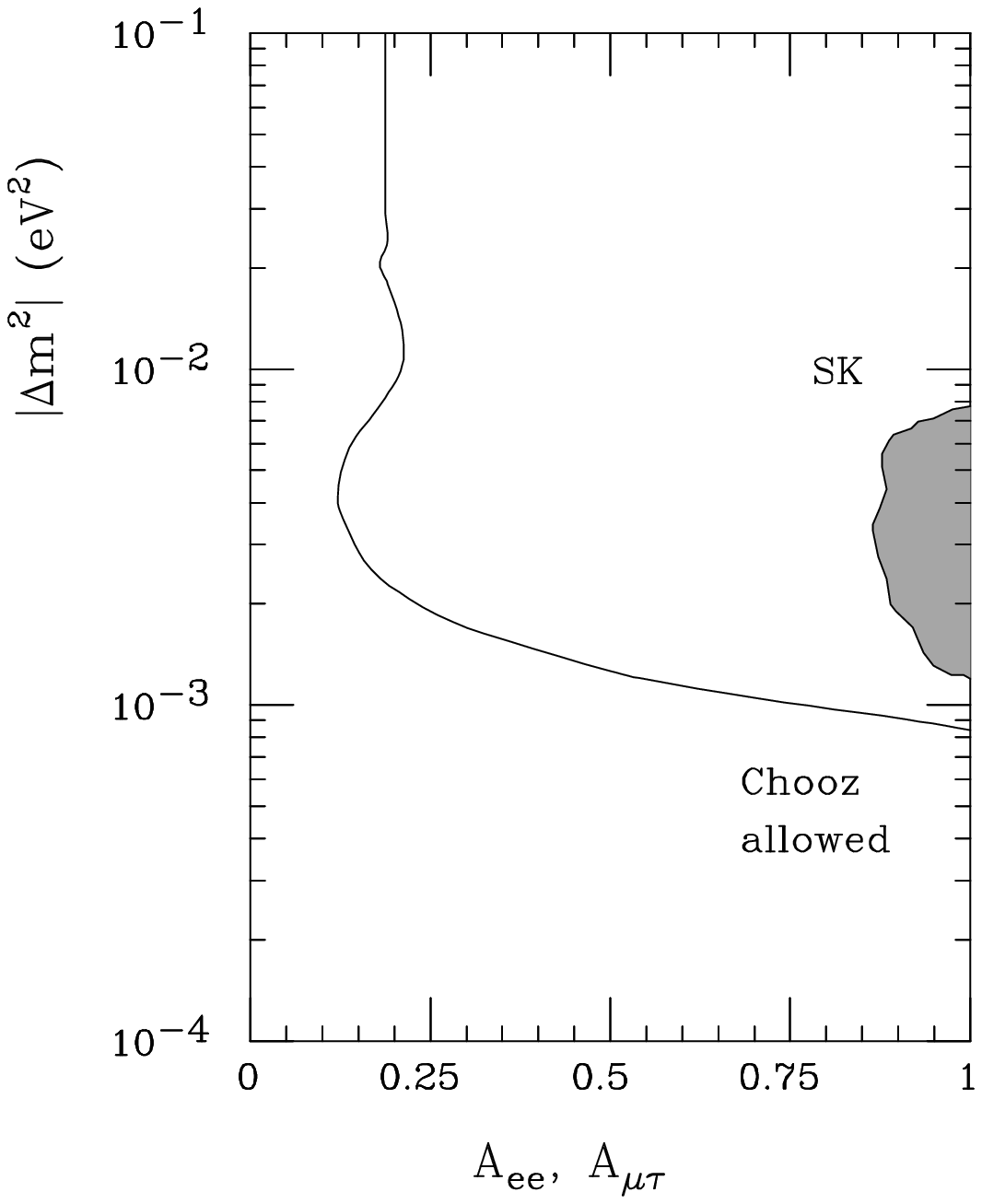,height=14cm}}
\caption {Limits on     $\nu_\mu \to \nu_\tau$ 
and $\nu_e \to \nu_e$  transitions 
obtained  by the Super--Kamiokande
and   Chooz experiments.
\label{fig:limit}
}
\end{figure}

The   available  information on  the  oscillation probabilities
in  the region of  interest  are  summarized in 
fig.~\ref{fig:limit}  where we  show the allowed   regions 
obtained  by the Super--Kamiokande  (shaded area)
and  the Chooz  experiment.
The  curves  are   taken  from \cite{SK-new}  and
\cite{Chooz}  where they were  obtained       in the framework of
of  a  two  flavor  oscillation  analysis
(we have only relabelled the  $x$--axis  
as  $A_{ee}$  for Chooz limit and 
$A_{\mu\tau}$  for the  SK  result).

We  want to translate   these  results   into 
constraints on  $p_{e 3}$, $p_{\mu 3}$ and $p_{\tau 3}$.
In the  Chooz  analysis the 
electron neutrino survival probability
is   described with  the 
form $P_{\nu_e \to \nu_e } = 1 - A_{ee} ~\sin^2 [\Delta m^2 /4 E_\nu]$,
and an allowed  region  is  determined  for the  parameters
$\Delta m^2$  and $A_{ee}$. The  form of the probability
is  identical  to  the one    obtained in 
the framework  of the one mass scale  approximation, and 
using $A_{ee} = 4 p_{e 3} \,(1 - p_{e 3})$   for  any  $\Delta m^2$ the
allowed  interval in $A_{ee}$ can  be  translated in an allowed interval
on $p_{e 3}$   (composed of two  disconnected  sub--intervals).

The SK analysis  is  valid  only in the  framework
of two  flavor  ($\nu_\mu \leftrightarrow \nu_\tau$) oscillations, 
since it is  assumed  that the $\nu_e$  and  $\overline{\nu}_e$ fluxes are not
modified.
In the framework of the one mass scale approximation,  
this  corresponds to the  limit $p_{e 3} = 0$.  
In this  limit, using $A_{\mu\tau} = 4 \;p_{\mu 3} \; p_{\tau 3}$,
the allowed region in $A_{\mu \tau}$  can be  translated  in 
allowed intervals for $p_{\mu 3}$ and $p_{\tau 3}$.
Since the Chooz  results  tell us 
that $p_{e 3}$  is  small 
we  can consider  these intervals  as  a reasonable approximation.
Taking into account the possibility of a non vanishing
$p_{e 3}$ the  allowed  interval for $p_{\mu 3}$ (or $p_{\tau 3}$ 
is  slightly enlarged (for a  detailed discussion see \cite{Fogli}).

\begin{figure} [t]
\centerline{\psfig{figure=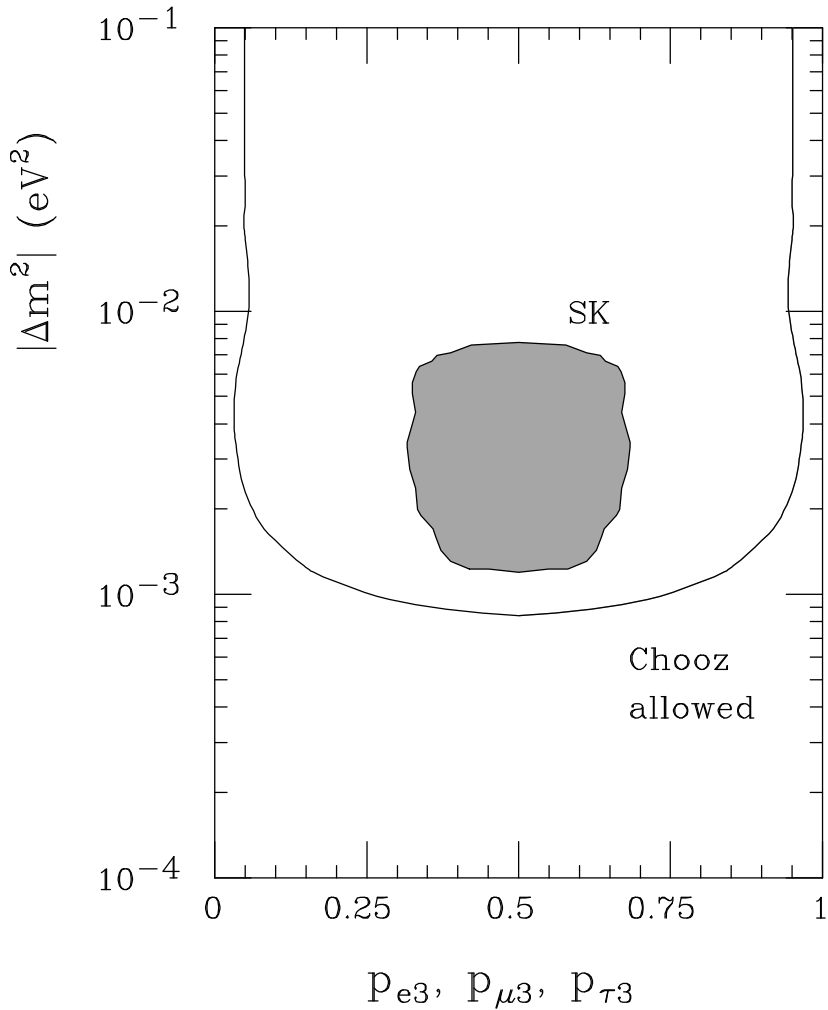,height=14cm}}
\caption {Limits  on the flavor  content
$p_{\alpha 3} = |\langle \nu_\alpha | \nu_3\rangle |^2$
of the  neutrino  state $|\nu_3\rangle$  
obtained from the  results of 
the SK
and   Chooz experiment. 
The  SK results    give  an allowed interval for
$p_{\mu 3}$ and $p_{\tau 3}$;
the  Chooz results   give an  allowed  interval 
for $p_{e 3}$.
\label{fig:p3}
}
\end{figure}

The results  are  shown  in fig.~\ref{fig:p3}.
As  an illustration of how  to read the  figure
we  can  consider   the  value
$|\Delta m^2 | = 3 \times 10^{-3}$~eV$^2$.  For this  value
of the squared mass difference   the Chooz  upper limit 
($A_{ee} \le 0.13$)  tells us that the 
state $|\nu_3\rangle$ is  either  a quasi pure  electron
neutrino state or  contains  only a  small
overlap with $|\nu_e\rangle$: 
\begin {equation}
p_{e 3} \le 0.033 ~~~ {\rm or} ~~~~
p_{e 3} \ge 0.966 
\end{equation}
The SK  results   tell us  that there  are
oscillations  between   $\nu_\mu$ and $\nu_\tau$ 
and the amplitude  of the  oscillations is close  to unity
($A_{\mu\tau} \ge 0.86$).  This can be  translated   as:
\begin {equation}
0.32 \le p_{\mu 3}, p_{\tau 3}  \le 0.68
\end{equation}
Of course  the three probabilities  are constrained to satisfy
$p_{e3} + p_{\mu 3 } + p_{\tau  3} = 1$, 
therefore only the `small $e$--flavor'
interpretation of  the Chooz  data remains  acceptable, and
a large (small) $p_{\mu 3}$  implies a small (large) $p_{\tau 3}$.
We note that the best fit of Super--Kamiokande 
($\sin^2 2 \theta_{\mu\tau} = 1$)  corresponds
 to  $p_{\mu 3} = p_{\tau 3} = {1\over 2}$ and
$p_{e 3} = 0$.

A central  goal  of future experiments 
will be to  measure   more accurately the flavor content  of the
neutrino state $|\nu_3\rangle$, and in particular   to  measure 
(or  put a more stringent  limit)
to  the overlap   between   the $|\nu_3\rangle$ and
$|\nu_e\rangle$ states.

We  note  that,  in analogy with  the  charged  lepton  and  
quark masses, it is  `natural' to expect that the state
$|\nu_3\rangle$, that is  experimentally determined to be 
a combination   with approximately equal  weights 
of muon and  tau  neutrinos   and a small  (or  vanishing) $|\nu_e\rangle$
component, is the heaviest  neutrino.
This  however is  an  assumption that could  be  false,
and that  should be experimentally   verified.
No  oscillation  experiment in vacuum   can solve
this    ambiguity,  since the mass dependent oscillating  term
$\sin^2[\Delta m^2 L/(4 E_\nu)]$ is  invariant for a
change of  sign in $\Delta m^2$.  
If  however  neutrinos  are  propagating in matter,    measurements of
flavor transitions  can determine  the  sign of $\Delta m^2$, that
is  determine if the state $|\nu_3\rangle$ is the
heaviest or  lightest  neutrino.

\section {Matter  effects}
When neutrinos  propagate in  matter
the  interactions    with the  medium   results  in  a 
a  flavor  dependent   effective potential  $V_\alpha$ \cite{MSW}.
The  difference in  effective  potential
between   electron neutrinos  and (tau) muon  (anti)--neutrinos is:
\begin{equation}
V = V_e  - V_\mu = V_e - V_\tau = \pm\sqrt{2} \;G_F~ n_e
\end {equation}
where $n_e$ is the electron density of  the medium,
$G_F$ is  the Fermi constant  and 
the plus  (minus)  sign applies to 
neutrinos  (anti--neutrinos).
Using the relation 
\begin{equation}
E_\nu = \sqrt{p^2 + m^2} + V \simeq  p + { m^2 \over 2 p}  + V
\end{equation}
the potential can be  considered  as
a contribution  $\delta m^2 (\nu_e)  = 2 E_\nu V$
to the effective  squared mass  of $\nu_e$'s.
The neutrino effective  squared  mass 
eigenvalues  $M_j^2$  and    the  mixing   matrix  in  matter  $U_m$ can 
calculated  solving the equation:
\begin {equation}
U~\left [
 \begin{array} {c c c}
  m_1^2  & 0      & 0 \\
  0      & m_2^2  & 0 \\
  0      & 0      & m_3^2  
 \end {array}
\right ]
U^T + 
\left [
 \begin{array} {c c c}
  2 \, V \, E_\nu  & 0      & 0 \\
  0   & 0 & 0 \\
  0   & 0 & 0  
 \end {array}
\right ]
 = 
U_m ~\left [
 \begin{array} {c c c}
  M_1^2  & 0      & 0 \\
  0      & M_2^2  & 0 \\
  0      & 0      & M_3^2  
 \end {array}
\right ]
~U_m^T
\label{eq:diagonalize}
\end{equation}
The  columns   of the matrix  $U_{m}$  give the    flavor  components
of  a  new set of    `propagation  eigenvectors' \{$|\nu_{1 m}\rangle$, 
$|\nu_{3 m}\rangle$, $|\nu_{3 m}\rangle$\}    with   well  defined
effective mass in matter.

The  matrix  diagonalization problem  of 
equation (\ref{eq:diagonalize})  is  very simple    under the
assumption $m_1^2 = m_2^2$  and has  been discussed before  by several authors
\cite{Pantaleone,Fogli,ADLS}.
The neutrino state
$|\nu_1\rangle$   has been chosen (see previous section)
as  having  no electron  flavor  component, and therefore  
is  decoupled from all  matter effects,   and the problem  is 
equivalent to the well  known  case of
two  flavor  mixing. 
The  effective mass eigenvalues  and the mixing  matrix can  be
written   as  a  function of the  adimensional quantity $x$:
\begin  {equation}
x = {2 V E_\nu \over \Delta m^2 } =
\pm {2 \sqrt {2} \;G_F \; n_e\, E_\nu \over  \Delta m^2} 
\simeq  \pm 0.076 \; {\rho(g\,cm^{-3}) ~E_\nu(GeV) \over
\Delta m^2 (10^{-3}~eV^2) }
\label{eq:xm}
\end{equation}
the plus (minus)  sign  applies  to  $\nu$'s ($\overline{\nu}$'s).
For the  numerical  estimate in (\ref{eq:xm})  we have also assumed
an electron fraction $Y_e = 0.5$.

The  effective squared mass   eigenvalues  are:
\begin {eqnarray}
M_1^2 & = & m_1^2  \nonumber \\
M_2^2 & = & m_1^2  + \Delta m^2 ~{1\over 2}  \left [
1 + x - \sqrt{ \sin^2 2 \theta  + (x - \cos 2 \theta)^2 } \right ]
\label{eq:masses}
 \\
M_3^2 & = & m_1^2  + \Delta m^2 ~{1\over 2} \left [
1 + x  + \sqrt{ \sin^2 2 \theta  + (x - \cos 2 \theta)^2 }
\right ] \nonumber
\end{eqnarray}

The  mixing  matrix in matter $U_m$  has the  same  form 
as  in the vacuum  case:
\begin {equation}
U_m = \left [
 \begin{array} {c c c}
  0 &  \cos \theta_m(x) & \sin \theta_m(x)  \\
 \cos \varphi & -\sin\theta_m (x)  \sin \varphi 
               & \cos \theta_m (x) \sin \varphi \\
 -\sin \varphi  & -\sin \theta_m (x) \cos \varphi 
               & \cos \theta_m (x) \cos \varphi 
 \end {array}
\right]
\end{equation}
but the angle $\theta_m$  is  a  function of the parameter $x$:
\begin {equation}
\sin \theta_m (x) = 
{ (x - \cos 2 \theta) + \sqrt{ \sin^2 2 \theta  +  (x - \cos 2 \theta)^2 }  
\over
\{
[(x - \cos 2 \theta) + \sqrt{ \sin^2 2 \theta  +  (x - \cos 2 \theta)^2} ]^2 +
 \sin^2  2 \theta \}^{ {1\over 2} }  }
\label {eq:st}
\end{equation}
The  resulting $\sin^2 2  \theta_m (x)$    has  the well  known expression:
\begin {equation}
\sin^2 2  \theta_m (x) = 
{ \sin^2 2 \theta \over
\sin^2 2 \theta  +  (x - \cos 2 \theta)^2 }
\label {eq:st2}
\end{equation}

Note that  we have    given  explicitely the 
three mass  eigenvalues   (and  not only the difference
$M_3^2 - M_2^2$), and also
$\sin \theta_m$   (and  not only 
$\sin^2 2 \theta_m$), because   these  quantities  are  needed 
to  compute   3--flavor oscillations, as  we  will discuss 
in the following.

\begin{figure} [t]
\centerline{\psfig{figure=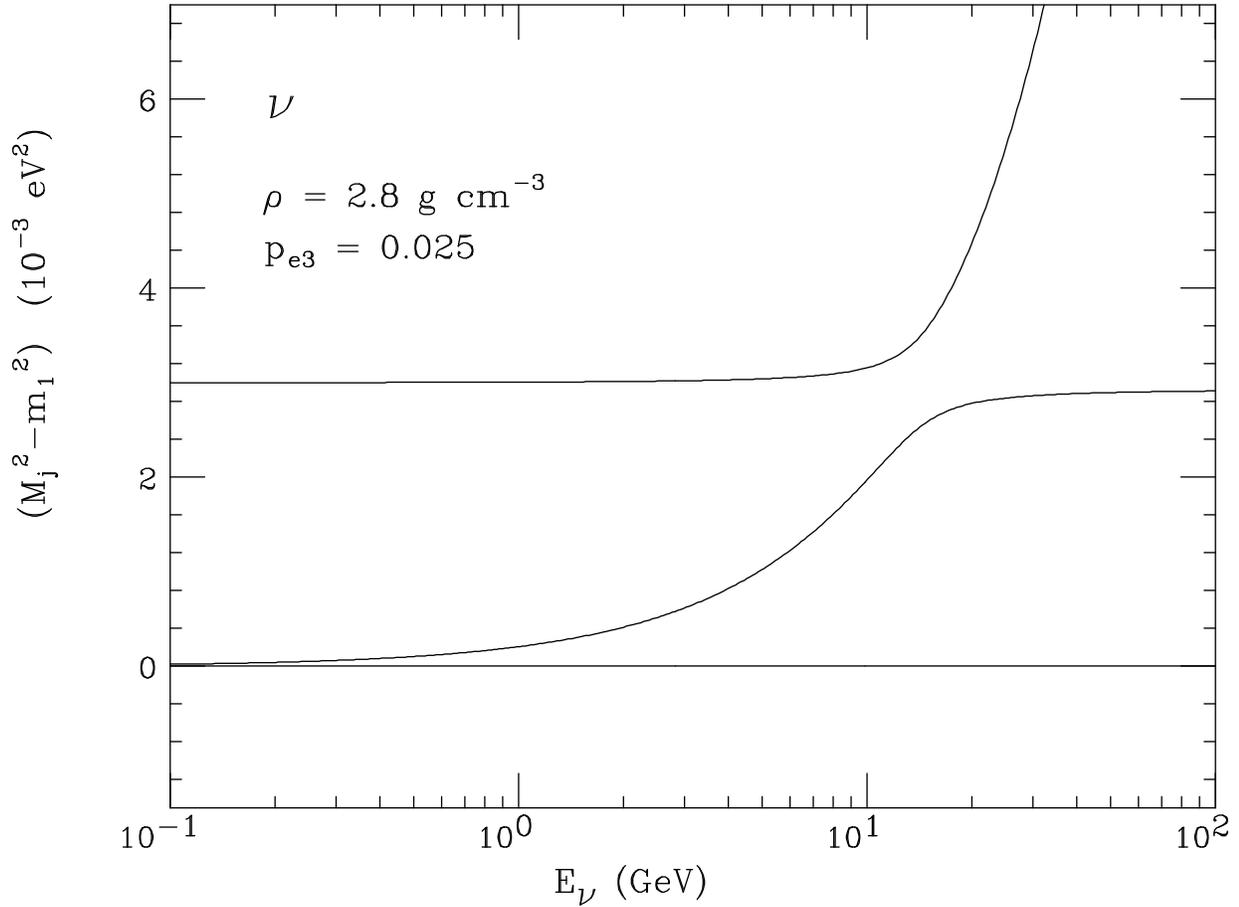,height=12.5cm}}
\caption { Effective mass eigenvalues 
for neutrinos propagating  in matter with density
$\rho = 2.8$~g~cm$^{-3}$  (and electron  
fraction $Y_e = n_e/(n_p + n_n) = 1/2$) plotted 
as  a  function of  $E_\nu$.
The    neutrino masses are   $m_1^2 = m_2^2$,
and $m_3^2  = m_1^2 + 3 \cdot 10^{-3}$~eV$^2$.
The  neutrino state  $|\nu_3\rangle$  as  a probability 
$p_{e 3} = \sin^2 \theta = 0.025$ of having   electron  flavor.
\label{fig:dm_nu}
}
\end{figure}

\begin{figure} [t]
\centerline{\psfig{figure=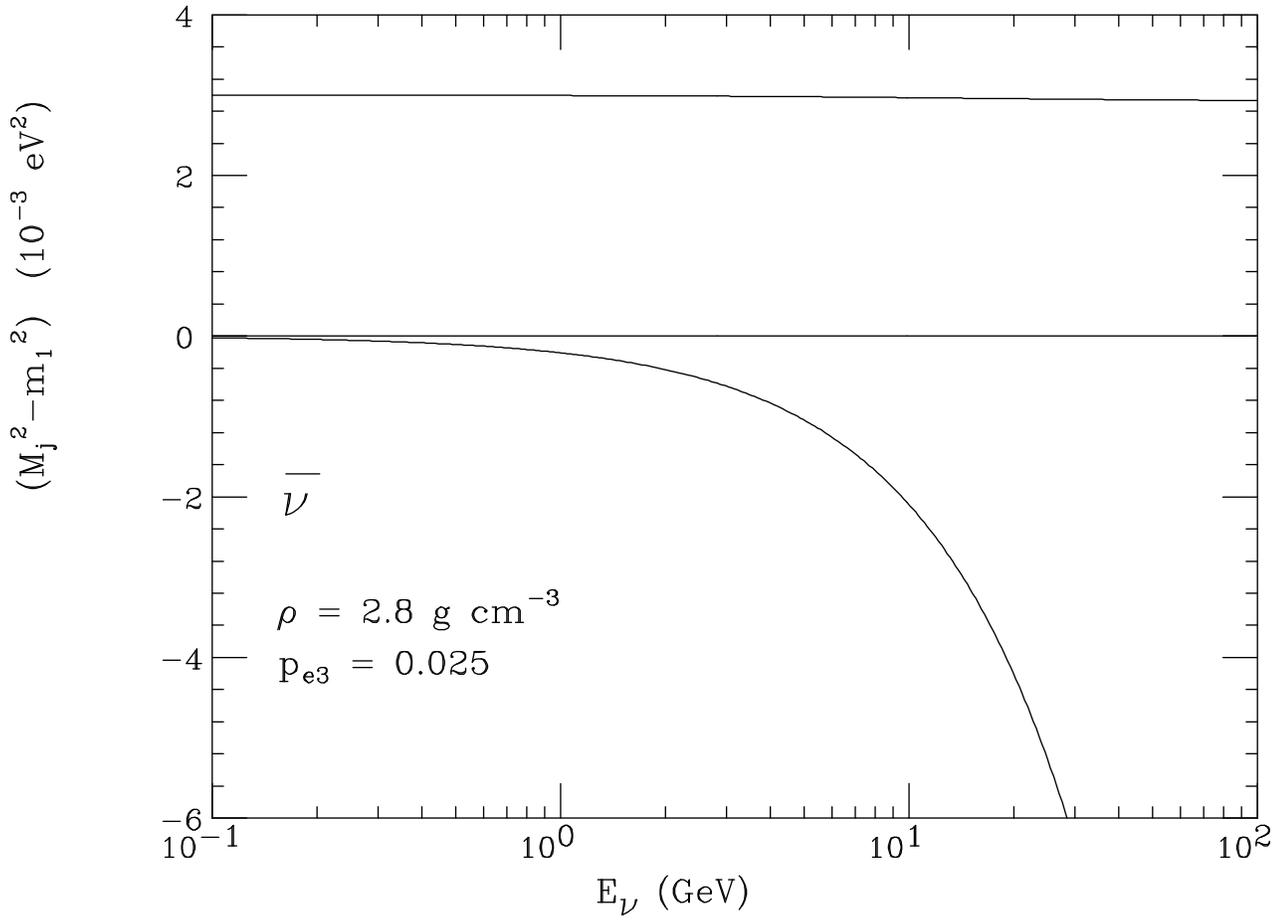,height=12.5cm}}
\caption {Effective mass eigenvalues 
of anti--neutrinos. 
propagating  in matter with density
$\rho = 2.8$~g~cm$^{-3}$.
The neutrino masses and  mixing are the  same as in fig.~1.
\label{fig:dm_nubar}
}
\end{figure}

\begin{figure} [t]
\centerline{\psfig{figure=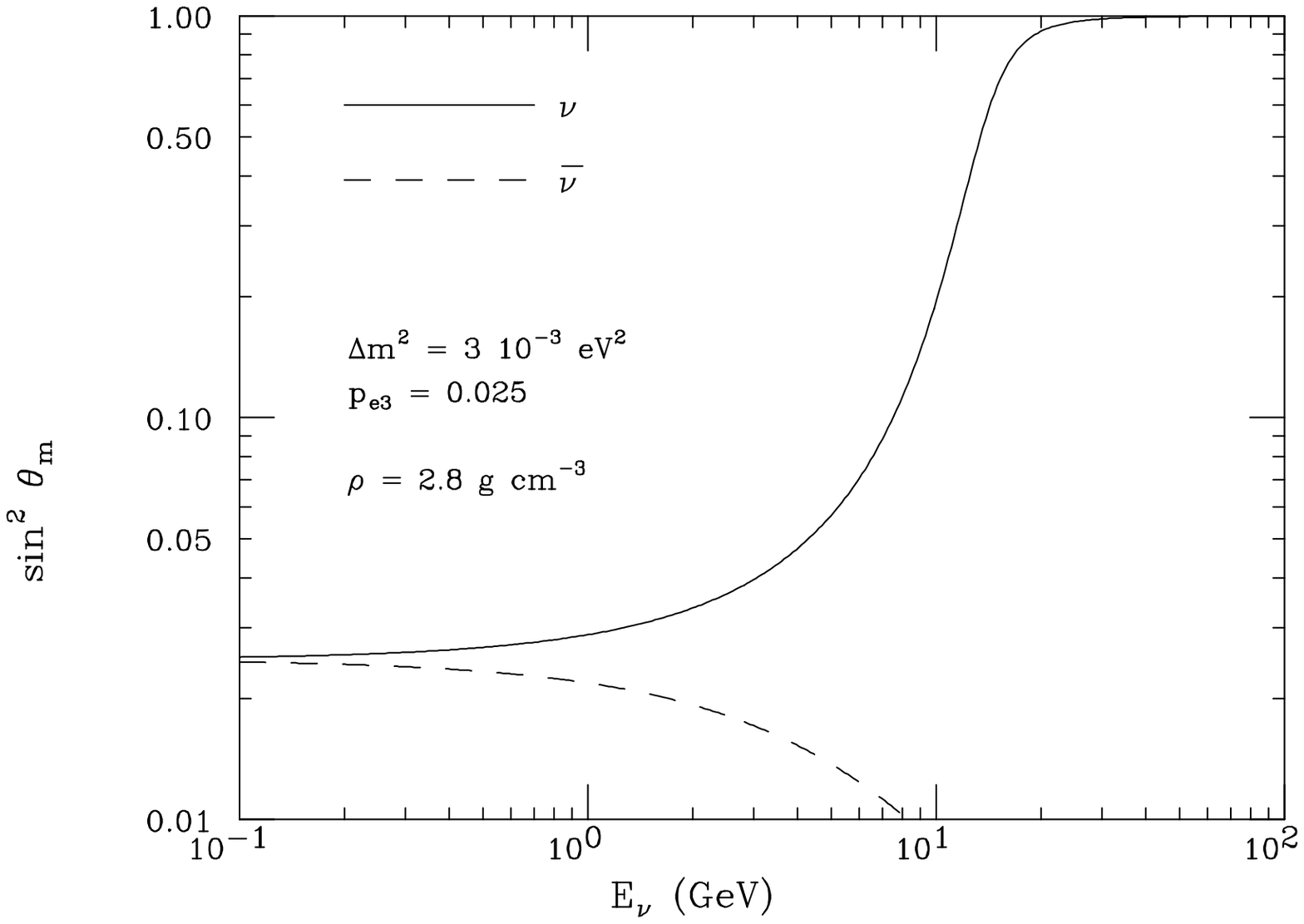,height=12.5cm}}
\caption {Value of $\sin^2 \theta_m$  
plotted  as  a  function of $E_\nu$  for  neutrinos (solid line)
and antineutrinos  (dashed line) propagating  in matter  with 
density $\rho = 2.8$~g~cm$^{-3}$.
In vacuum $p_{e 3} = \sin^2 \theta = 0.025$.
The neutrino masses  are $m_1 = m_2$,
$m_3^2  = m_1^2 + 3 \cdot 10^{-3}$~eV$^2$.
\label{fig:stm}
}
\end{figure}

\begin{figure} [t]
\centerline{\psfig{figure=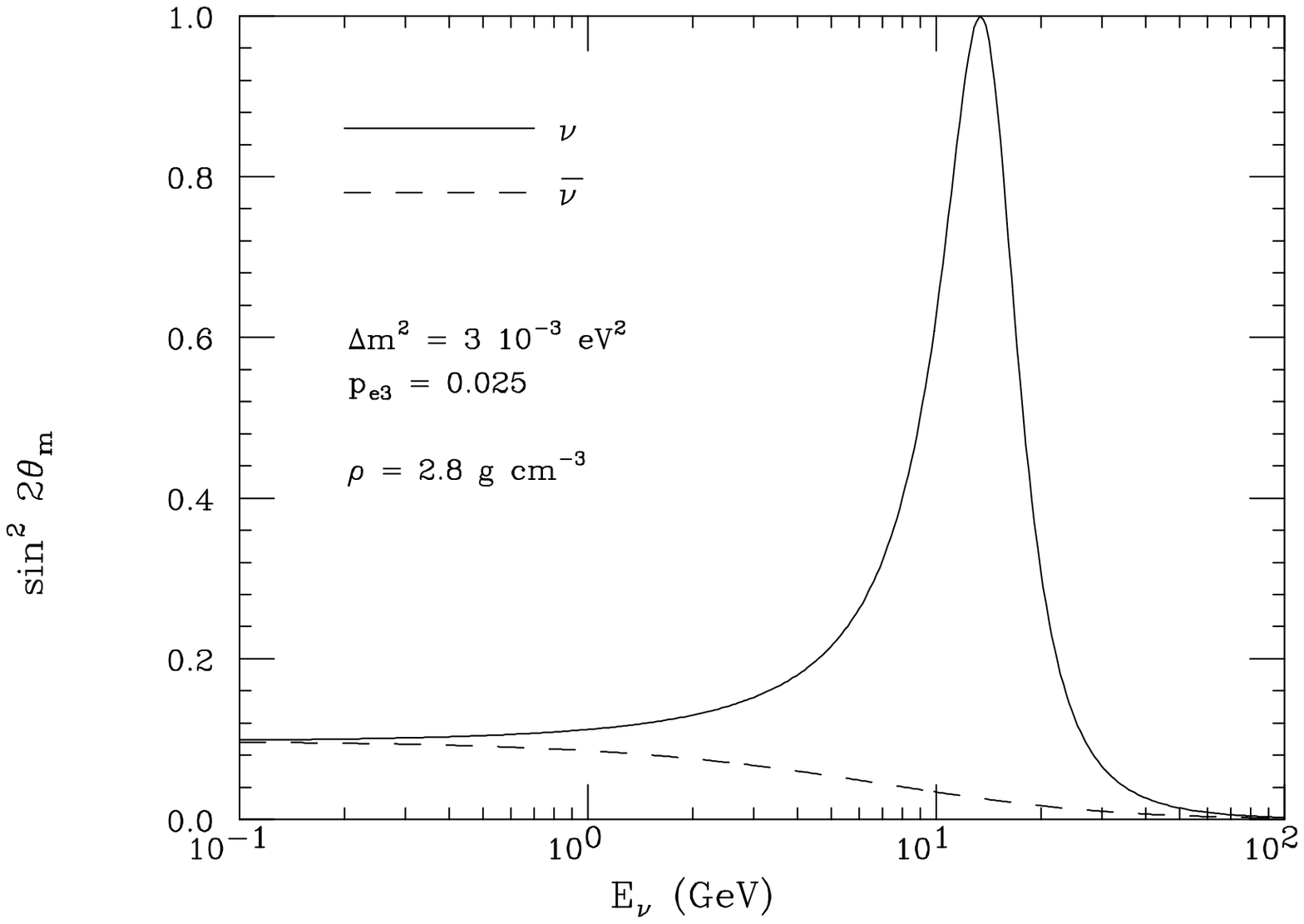,height=12.5cm}}
\caption {Value of $\sin^2 2 \theta_m$  
plotted  as  a  function of $E_\nu$  for  neutrinos (solid line)
and antineutrinos  (dashed line) propagating  in matter  with 
density $\rho = 2.8$~g~cm$^{-3}$.
In vacuum $p_{e 3} = \sin^2 \theta = 0.025$
($\sin^2 2 \theta = 0.0975$).
The neutrino masses  are $m_1 = m_2$,
$m_3^2  = m_1^2 + 3 \cdot 10^{-3}$~eV$^2$.
\label{fig:st2m}
}
\end{figure}

An illustration of the  effect of  matter  on the   masses 
and  mixing  of  the neutrinos  is  shown  in figures 3, 4, 5 and 6.
For these  figures  we have  assumed 
$\Delta m^2 = 3 \cdot 10^{-3}$~eV$^2$,
a probability $p_{e 3}=  |U_{e 3}|^2 \equiv \sin^2 \theta = 0.025$ 
(the specification of $U_{\mu 3}$   and   $U_{\tau 3}$ is   not  necessary)
and that the    neutrinos (or anti--neutrinos)  propagate in matter
of  constant density 
$\rho = 2.8$~g~cm$^{-3}$  (with electron  
fraction $Y_e = 1/2$).  This  last 
condition is  a good  approximation  
for neutrinos  in LBL  beams  that travel
few kilometers  below the  Earth's surface.
In  fig.~\ref{fig:dm_nu}   we show 
the neutrino effective mass
eigenvalues $M_j^2$    (equation \ref{eq:masses})  
plotted as  a function  of 
$E_\nu$. 
Fig.~\ref{fig:dm_nubar} is the same   but for  antineutrinos.
In  fig.~\ref{fig:stm}   (fig.~\ref{fig:st2m})  we show 
the values of  $\sin^2 \theta_m$   ($\sin^2 2 \theta_m$)
(again  plotted as a function of $E_\nu$),
the solid (dashed)  lines   refer to   $\nu$'s ($\overline{\nu}$'s).

Note that  in  matter  the propagation  eigenvectors 
 $|\nu_{1m}\rangle \equiv |\nu_1\rangle$ 
and  $|\nu_{2m}\rangle$  are
not anymore  degenerate,  since the  effective  mass  of 
$|\nu_1\rangle$   can be considered as   constant while the effective
squared mass  of the    state  $|\nu_{2m}\rangle$ 
($|\overline{\nu}_{2m}\rangle$ )  increases  (decreases)   with  increasing
$n_e\,E_\nu$.

Inspecting the equations  that describe  the effective masses
and  mixing in  matter,  we can  notice that there are
some  special   conditions:
\begin {enumerate}
\item The case $x = 0$ is  simply the  vacuum case:
$M_j^2(0) = m_j^2$ and  $\theta_m(0) = \theta$.
\item   The case  $x  \to +\infty$    corresponds  to the limit of
very large   density  or  very large $E_\nu$ 
for  neutrinos  (anti--neutrinos)  if 
$\Delta m^2 > 0$  ($\Delta m^2 < 0$).
In this  situation  the state  $|\nu_{3 m}\rangle$  becomes
a  pure $|\nu_e\rangle$ ($\theta_m \to \pi/2$) 
with a  very large  effective mass   $M_{3}^2 \to \infty$, and
decouples  from the  oscillations;
the state $|\nu_{2 m}\rangle$  asymptotically  has the effective mass
 $M_{2}^2 \to m_1^2 + \Delta m^2\, \cos^2 \theta$.
Note  that in this  condition 
of very strong matter effects, the oscillations involving 
$\nu_e$  are  completely suppressed, but 
$\nu_\mu \leftrightarrow  \nu_\tau$  oscillations 
do occur with a    maximum  probability $4 \sin^2 \varphi \cos^2 \varphi$
larger  than  the vacuum  one by  a factor  $(\cos^2 \theta)^{-1}$,
and  an effective  $\Delta m^2$  smaller   than the vacuum value
by a  factor  $\cos^2 \theta$.

\item   The case  $x  \to -\infty$    corresponds  to the limit of
very large   density, or  very large $E_\nu$
  for  anti--neutrinos  (neutrinos)  if 
$\Delta m^2 > 0$  ($\Delta m^2 < 0$).
In this  situation  the state  $|\nu_{2 m}\rangle$  becomes
a  pure $|\nu_e\rangle$ ($\theta_m  \to 0$) 
with effective mass   $M_{3}^2 \to -\infty$;
the state $|\nu_{3 m}\rangle$  asymptotically   gets  the  mass
$M_{3}^2 \to m_1^2 + \Delta m^2 \,\cos^2 \theta$.
This  situation  is  experimentally undistinguishable  from
the  previous  case.    
  
\item  Finally we have the special case:
\begin {equation}
x = \cos 2 \theta = 1 - 2 \,p_{e3}
\end{equation}
This case can  only happen  for  neutrinos  (anti--neutrinos)  if
$\Delta m^2 > 0$ ($\Delta m^2 < 0$).  It corresponds to the 
celebrated MSW  \cite {MSW} resonance.  
For a  fixed  electron  density, the  resonance  happens
at  a  value  of  the   neutrino  energy:
\begin{equation}
E_\nu^{res}  =
 {\Delta m^2 \;\cos 2 \theta \over 2 \,V } =  4.7 \;\cos 2 \theta ~ 
\left ( {|\Delta m^2 | \over 10^{-3} ~{\rm eV}^2 } \right )~
\left ( {2.8 ~{\rm g~cm}^3 \over \rho}  \right )
~\left ( {0.5 \over Y_e}  \right )
 ~{\rm GeV}
\end{equation}
At the  resonance the mixing angle in matter
is  $\theta_m = 45^\circ$,
the probability  $P(\nu_e\to\nu_e)$  
oscillates  with  amplitude  $\sin^2 2 \theta_{m} = 1$,    while 
the probabilities   $P(\nu_e\to\nu_\mu)$  
and   $P(\nu_e\to\nu_\tau)$  
oscillate  with amplitudes  $\sin^2 \varphi$  
and  $\cos^2 \varphi$. 
\end{enumerate}

\section {Propagation in matter  of constant  density}

\begin{figure} [t]
\centerline{\psfig{figure=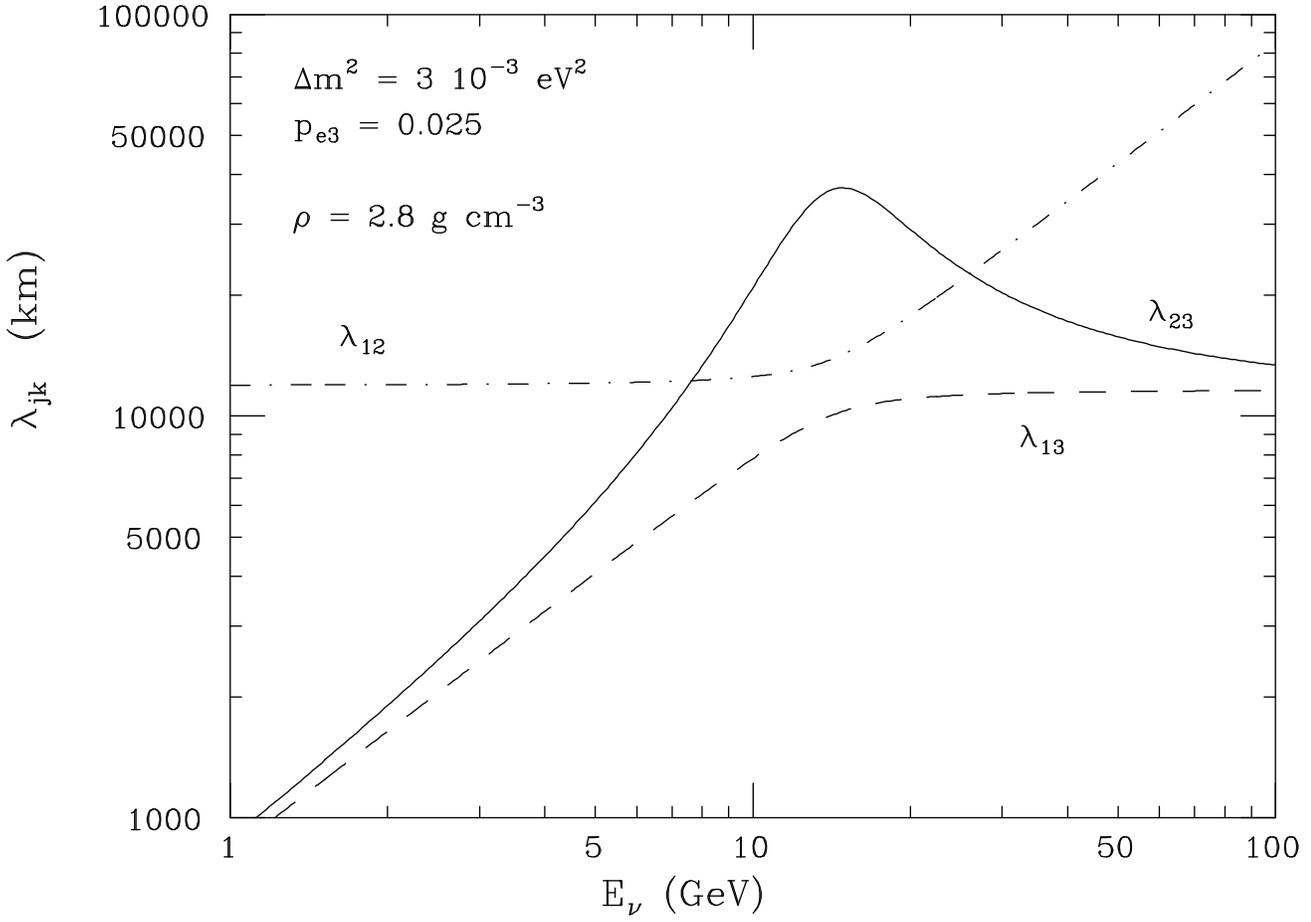,height=12.5cm}}
\caption {     Oscillation  lengths   $\lambda_{jk}  
= 4\pi E_\nu/|M_j^2 - M_k^2|$  plotted as  a function of
$E_\nu$ for neutrinos  traveling in matter
of density  $\rho= 2.8$~g~cm$^{-3}$.
The neutrino masses  are $m_1 = m_2$,
$m_3^2  = m_1^2 + 3 \cdot 10^{-3}$~eV$^2$,
the electron flavor  content  of the state  $|\nu_3\rangle$ is
$p_{e 3} = \sin^2 \theta = 0.025$.
\label{fig:l3}
}
\end{figure}

For  neutrinos  propagating in   matter of a constant  density the
oscillation probabilities    can be  written  in general  as
the   superposition of three  oscillating terms
corresponding to the three effective squared  mass differences:
\begin{equation}
P_{\nu_\alpha \to \nu_\beta} (L,E_\nu; ~x) 
= \sum_{j < k}  A_{\alpha \beta}^{jk} (x) 
\sin^2 \left [ \pi {L \over \lambda_{jk} (x, E_\nu) } \right ].
\label{eq:prob3}
\end {equation}
The amplitudes of the oscillatig terms are given by:
\begin{equation}
A_{\alpha \beta}^{jk} (x)  = -4 \; 
U^m_{\alpha j} \, U^m_{\beta j}
\, U^m_{\alpha k} \, U^m_{\beta k}
\label{eq:ampl3}
\end {equation}
The  three  oscillation lengths   are:
\begin{equation}
\lambda_{jk} (E_\nu, ~x)  =  {4\pi E_\nu \over |M_j^2 - M_k^2| }
\end {equation}
and   satisfy  the relation:
\begin{equation}
\lambda_{13}^{-1} - \lambda_{12}^{-1} - \lambda_{23}^{-1} = 0
\end {equation}
They can be written  explicitely as
 a  function of the   parameter $x$: 
\begin {eqnarray}
\lambda_{12} (E_\nu, ~x) & =  & \lambda_0(E_\nu) ~\left [ {1\over 2} 
\left ( 1 + x - \sqrt {F(x)} \right ) \right ]^{ -1 } 
\label{eq:lam12}
\\
\lambda_{13} (E_\nu, ~x) & =  & \lambda_0 (E_\nu) ~\left [ {1\over 2} 
\left ( 1 + x + \sqrt {F(x)} \right ) \right ]^{ -1 } 
\label{eq:lam13}
\\
\lambda_{23} (E_\nu, ~x) & =  & 
\lambda_0 (E_\nu) ~\left [ \sqrt {F(x)} \right ]^{-1}
\label{eq:lam23}
\end{eqnarray}
where  we  have introduced the definition
\begin{equation}
F(x) = \sin^2 2 \theta + ( x - \cos 2 \theta)^2
\label{eq:Fx}
\end{equation}
In  fig.~\ref{fig:l3}   we  show as an example a plot of 
 the  oscillation  lengths   $\lambda_{jk}$
calculated as a function of 
the neutrino  energy for the  same  values of  the   neutrino  masses,
and  mixing and the same  density of the medium as  in the previous  figures.
Some  features  are  immediately  visible looking
at  fig.~\ref{fig:l3}:
\begin {enumerate}
\item  The presence of  matter  introduces  an energy
independent length:
\begin{equation}
\lambda_m(n_e) = {2 \pi \over |V|}  = 1.16 \cdot 10^4~{\rm km}~
\left ( { 1.69 \cdot 10^{24}~{\rm cm}^{-3} \over n_e} \right )
\end{equation}
\item  When  $x \to  0$ (low  $n_e ~E_\nu$) we  have
$\lambda_{12} \to  \lambda_m$,
$\lambda_{13},\lambda_{23} \to   \lambda_0$.
When 
$x \to  \infty$ (high   $n_e ~E_\nu$)  we  have
$\lambda_{12} \to  \lambda_0$,
$\lambda_{13},\lambda_{23} \to   \lambda_m$.
 
\item  The resonance condition can  also be 
expressed as $\lambda_0(E_\nu)  =  \lambda_m (n_e) \, \cos 2 \theta$.
At the resonance  the length 
$\lambda_{23}$   has  a maximum: $\lambda_{23} = \lambda_0/\sin 2 \theta$.
\end{enumerate}

For a fixed  value of $E_\nu$  
the oscillation  probabilities  involving  $\nu_e$  (or  $\overline{\nu}_e$'s)
have  a  simple  sinusoidal   dependence  on $L$  with 
oscillation  length $\lambda_{23}$.
This  can be understood  immediately  observing that
the  eigenstate  
$|\nu_1\rangle$     has  no    electron  flavor  component.
Formally we  have that     $U^{m}_{e 1} = 0$,   and    therefore 
from (\ref{eq:ampl3})  follows that:
\begin {equation}
A_{e\mu}^{12} = 
A_{e\mu}^{13} = 
A_{e\tau}^{12} = 
A_{e\tau}^{13} = 0
\end{equation}
therefore  the   transitions  
$\nu_e \leftrightarrow \nu_\mu$   and
$\nu_e \leftrightarrow \nu_\tau$     develop  with a  single
oscillation   length  $\lambda_{23}$.
The  probability
$P(\nu_\mu \leftrightarrow \nu_\tau)$, 
and therefore  also  $P(\nu_\mu \to \nu_\mu)$, have a more
complex functional  form  
because all the three  oscillating terms   in (\ref{eq:prob3}) 
are non  vanishing.
The  non  vanishing  probability  amplitudes
$A_{\alpha \beta}^{jk}$  can be   written  as:
\begin {eqnarray}
A_{e \mu}^{23}  (x)  & =  4 \sin^2 \theta_m \cos^2 \theta_m 
\sin^2 \varphi  
& = \sin^2 2 \theta_m \sin^2 \varphi   
\nonumber \\
A_{e \tau}^{23} (x)    & =  4 \sin^2 \theta_m \cos^2 \theta_m 
\cos^2 \varphi 
& = \sin^2 2 \theta_m \cos^2 \varphi  
\nonumber \\
A_{\mu \tau}^{23} (x)  & = - 4 \sin^2 \theta_m \cos^2 \theta_m 
\sin^2 \varphi  \cos^2 \varphi 
& = -\sin^2 2 \theta_m \sin^2 \varphi   \cos^2 \varphi 
\label {eq:ampl} \\
A_{\mu \tau}^{12} (x)  & =  4 \sin^2 \theta_m \sin^2 \varphi 
\cos^2 \varphi 
& ~ 
\nonumber \\
A_{\mu \tau}^{13}  (x) & =  4 \cos^2 \theta_m \sin^2 \varphi \cos^2 \varphi 
& ~ \nonumber
\end{eqnarray}

\subsection {Small  $L$  regime}
It is  interesting  to study the oscillation
probabilities  in matter in the limit of short neutrino pathlength $L$,
or  more  precisely in the limit where $L/\lambda_{jk} \ll 1$  for all
three oscillation lengths.
The  condition of `small $L$'   
is  going to be valid in the range $E_\nu \aprge 10$~GeV
for  the long--baseline  beams such as  the Fermilab to Minos
and  CERN to  Gran Sasso projects  ($L \simeq  730$~km)
if $|\Delta m^2|$ is in the range indicated  by
the atmospheric  neutrino data.

It the  limit   of small $L$ the   oscillation  probabilities in   matter
are  undistinguishable from the vacuum  case, even if the  
effective masses  and  mixing in matter  are    very  different from the 
vacuum  case.
This  can be  demonstrated  observing that 
when $L/\lambda_{jk}$ is small
the oscillating  terms  in  (\ref{eq:prob3})
can be approximated as:
\begin {equation}
\sin^2 \left (  \pi \;{L \over \lambda_{jk} } \right ) \simeq
\left ( \pi { L \over \lambda_{jk} } \right)^2
\label {eq:expansion}
\end{equation}
and the oscillation probabilities  take the form:
\begin {equation}
P_{\nu_\alpha \to \nu_\beta}  =
(\pi\, L)^2 \; \sum_{j<k}  {A^{jk}_{\alpha \beta} (x) \over
 \lambda_{jk}^2 (E_\nu, x) }
\label {eq:exp3}
\end{equation}
There are  now two  relations
between  the  amplitudes  $A_{\alpha\beta}^{jk}$ and  the  oscillations
lengths  $\lambda_{jk}$   that  result in a  cancellation of the 
$x$ dependence of the expression (\ref{eq:exp3}).
The  first  relation is  present also in the case of two
flavor mixing:
\begin {equation}
{\sin^2 2 \theta_m (x)\over \lambda_{23}^2(E_\nu, x) } = 
\left ( { \sin^2 2 \theta \over F(x)} \right )
~\left (
{\sqrt{F(x)} \over \lambda_0(E_\nu)} \right )^2  = {\sin^2 2 \theta 
\over \lambda_0^2 (E_\nu)}
\label {eq:rel1}
\end {equation}
where we have used   equations 
(\ref{eq:lam23})  and (\ref{eq:st2})   in the  first equality.
Similarly   it is  easy to prove that:
\begin {equation}
{ \sin^2 \theta_m \over  \lambda_{12}^2 } + 
{ \cos^2 \theta_m  \over  \lambda_{13}^2 }  =
{\cos^2  \theta \over \lambda_0^2 }.
\label {eq:rel2}
\end{equation}
From equation (\ref{eq:ampl})   we can see that 
$A_{\alpha\beta}^{23} \propto \sin^2 2 \theta_m$,
$A_{\alpha\beta}^{12} \propto \sin^2  \theta_m$ and 
$A_{\alpha\beta}^{13} \propto \cos^2  \theta_m$, and 
using  the relations (\ref{eq:rel1}) and (\ref{eq:rel2})
in (\ref{eq:exp3})   one  can  see  that 
for  $L/\lambda_{jk}$ small  the
matter  effects    have  no experimentally  detectable effects.

\begin{figure} [t]
\centerline{\psfig{figure=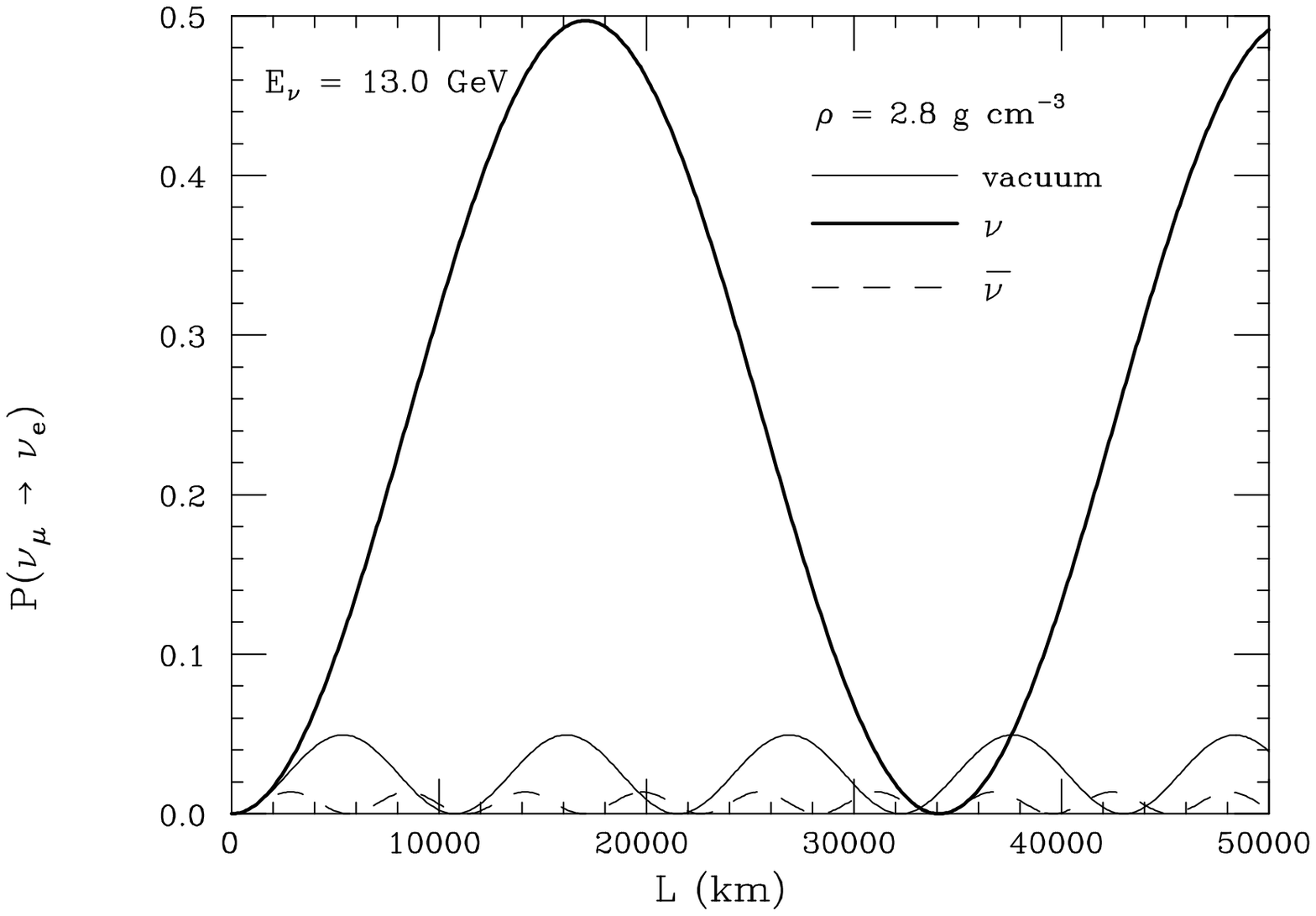,height=12.5cm}}
\caption {     Transition probability
$P(\nu_\mu \to \nu_e)$, plotted as  a function of
the distance $L$ for  neutrinos   with a fixed  energy
$E_\nu = 13.0$~GeV  traveling  in vacuum 
(solid  line)  or in matter  of
constant matter with density 
 $\rho = 2.8$~g~cm$^{-3}$
(thick solid   line for neutrinos,  dashed  line for antineutrinos).
The neutrino masses  are $m_1 = m_2$,
$m_3^2  = m_1^2 + 3 \cdot 10^{-3}$~eV$^2$.
The  mixing  matrix  in  vacuum is  determined
by $p_{e 3} = \sin^2 \theta=0.025$  and 
$p_{\mu 3} = p_{\tau 3}$ ($\sin^2 \varphi = 0.5$).
$E_\nu = 13.0$~GeV  is close to the resonance  energy for the density
and  parameter values   considered.
\label{fig:pe}
}

\end{figure}

\begin{figure} [t]
\centerline{\psfig{figure=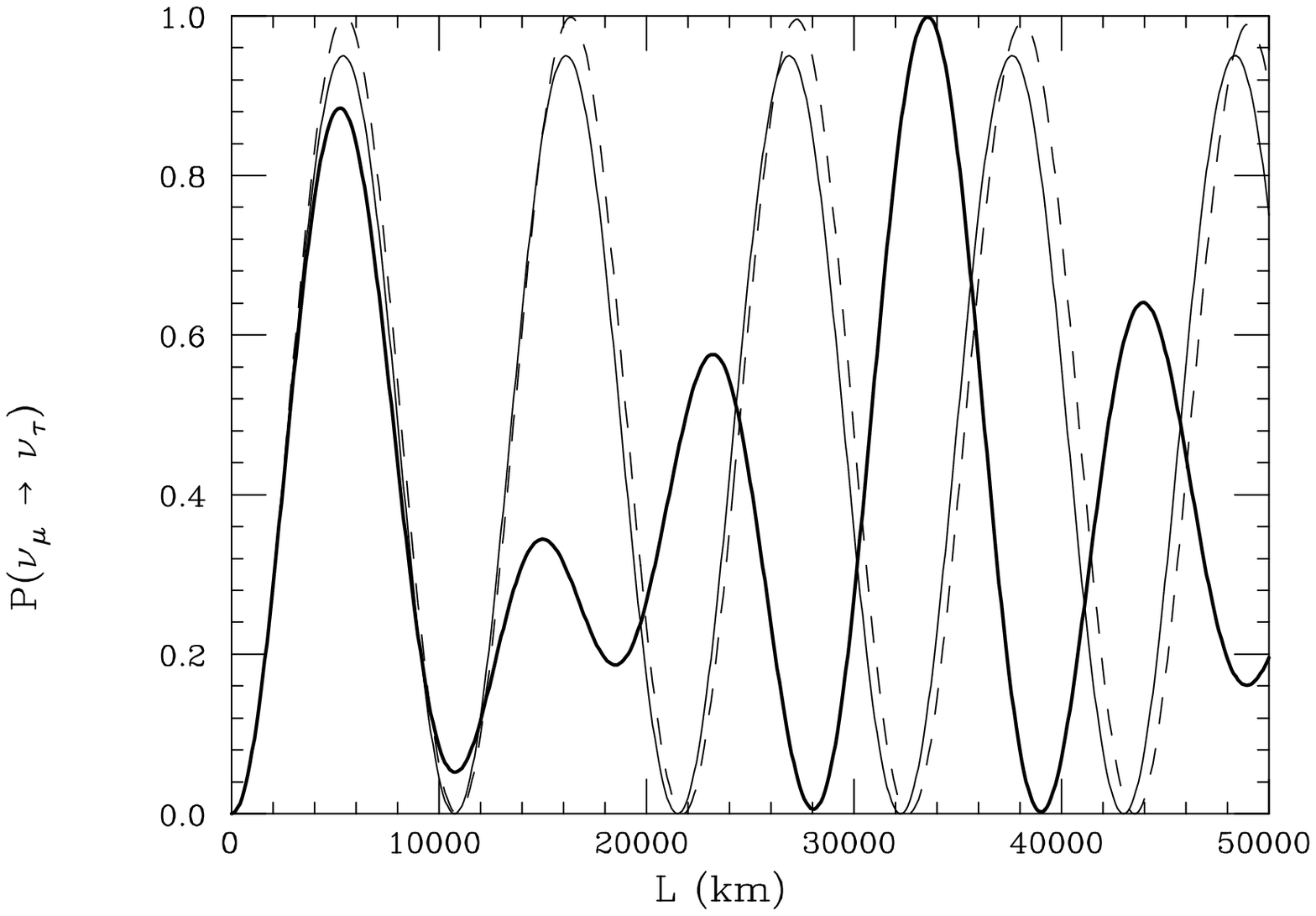,height=12.5cm}}
\caption {     Transition probability
$P(\nu_\mu \to \nu_\tau)$, plotted as  a function of
the distance $L$ for  neutrinos   with a fixed  energy
$E_\nu = 13.0$~GeV  traveling  in vacuum 
(thin solid  line)  or in matter  of
constant matter with density 
 $\rho = 2.8$~g~cm$^{-3}$
(thick solid   line for neutrinos,  dashed  line for antineutrinos).
The neutrino masses  are $m_1 = m_2$,
$m_3^2  = m_1^2 + 3 \cdot 10^{-3}$~eV$^2$.
The  mixing  matrix  in  vacuum is  determined
by $p_{e 3} = \sin^2 \theta=0.025$  and 
$p_{\mu 3} = p_{\tau 3}$  ($\sin^2 \varphi = 0.5$).
$E_\nu = 13.0$~GeV  is  close to the resonance  energy for the density
and  parameter  values considered.
\label{fig:ptau}
}

\end{figure}

To  illustrate  this  point in 
fig.~\ref{fig:pe}  and fig.~\ref{fig:ptau}
we show the transition probability
$P(\nu_\mu \to \nu_e)$ and 
$P(\nu_\mu \to \nu_\tau)$
plotted as  a function of
the distance $L$ for  neutrinos   and  anti--neutrinos of a fixed  energy
$E_\nu = 13.0$~GeV  traveling  in vacuum 
or in matter  of
constant matter with density 
 $\rho = 2.8$~g~cm$^{-3}$.
For this  figures we have assumed  
again for  illustration 
$\Delta m^2 = 3 \cdot 10^{-3}$~eV$^2$,
$p_{e 3} = \sin^2 \theta=0.025$  and $p_{\mu 3} = p_{\tau 3}$.
The energy $E_\nu = 13.0$~GeV  is close to  the neutrino resonance  
energy  (13.4~GeV)  for the density and  mass  matrix  considered.
Note that   the  maximum pathlength    considered in these plots
$L = 5\cdot 10^{4}$~km   is   approximately  four times longer than 
the Earth's  diameter and   therefore in most of the 
range of  pathlengths  considered 
the  condition   considered  cannot  be
realized  in practice.
Looking at the  figure it is immediately  evident that  
the oscillation properties of the neutrinos in  matter and in 
vacuum  are  dramatically different,  however    experiments   with 
a  baseline of $L= 250$ or 730~km 
will not measure  significant  departures from the vacuum  
oscillation probabilities.

\section {Oscillation probabilities for LBL experiments}

In a LBL experiment the neutrino pathlength  $L$ and 
the  electron  density  $n_e(y)$ along the
trajectory ($y\in [0,L]$ is a coordinate along the neutrino path)
can be  considered  as fixed, therefore what are experimentally accessible,
are  the oscillation probabilities  
$P_{\nu_{\alpha} \to \nu_{\beta}} (E_\nu)$ as function of the neutrino energy.
In the following we will also use the approximation to consider
$n_e$ as constant.

If $L$ and $n_e$ are  fixed, 
in the one mass scale approximation
for a long  baseline experiment  one   has to consider 
two  characteristic  neutrino  energies:
\begin {equation}
\varepsilon_0 = {|\Delta m^2| \;L \over 2 \pi}  = 0.59 ~|\Delta m^2_{-3}|
\left ( {L \over 730~{\rm km} } \right ) 
~{\rm GeV}
\end{equation}
and
\begin {equation}
\varepsilon_m = 
{|\Delta m^2|\over 2 V } =
{|\Delta m^2|\over 2 \sqrt{2} G_f n_e}  = 4.70
 ~|\Delta m^2_{-3}|
\left ( { 1.69 \cdot 10^{24} ~{\rm cm}^{-3} \over n_e} \right )
~{\rm GeV}
\end{equation}
where $\Delta m^2_{-3}$ is the value of  the neutrino
squared mass  difference in  units of $10^{-3}$~eV$^2$.
The  energy $\varepsilon_0$ corresponds to    the highest neutrino energy  
for  which the vacuum  transition probabilities  have a maximum
(or in different words it is the energy 
such that  $\lambda_0(\varepsilon_0) = 2L\,$).
The  energy $\varepsilon_m$ is the resonance  energy in the 
limit of small $p_{e 3}$:
\begin {equation}
E_\nu^{res} = \varepsilon_m \;\cos 2 \theta = \varepsilon_m \;(1- 2\,p_{e 3}),
\end{equation}
the  parameter  $x$    that  controls  the importance  of
the matter effect is given  by  $|x| = E_\nu/\varepsilon_m$,
and therefore for $E_\nu \ll \varepsilon_m$ the    presence of matters
has a negligible effect.
Both   characteristic   energies  are proportional to
$\Delta m^2$,  and the ratio  $\varepsilon_0/\varepsilon_m$ is  
independent from the neutrino mass matrix.
For the Fermilab  and  CERN projects  one has
$\varepsilon_0/\varepsilon_m = 0.125$, 
for the KEK to Kamioka (K2K) project  $\varepsilon_0/\varepsilon_m = 0.042$. 
The smalless of this  ratio
ensures that the  matter effects  are only a  correction  to the
oscillation probabilities.

\begin{figure} [t]
\centerline{\psfig{figure=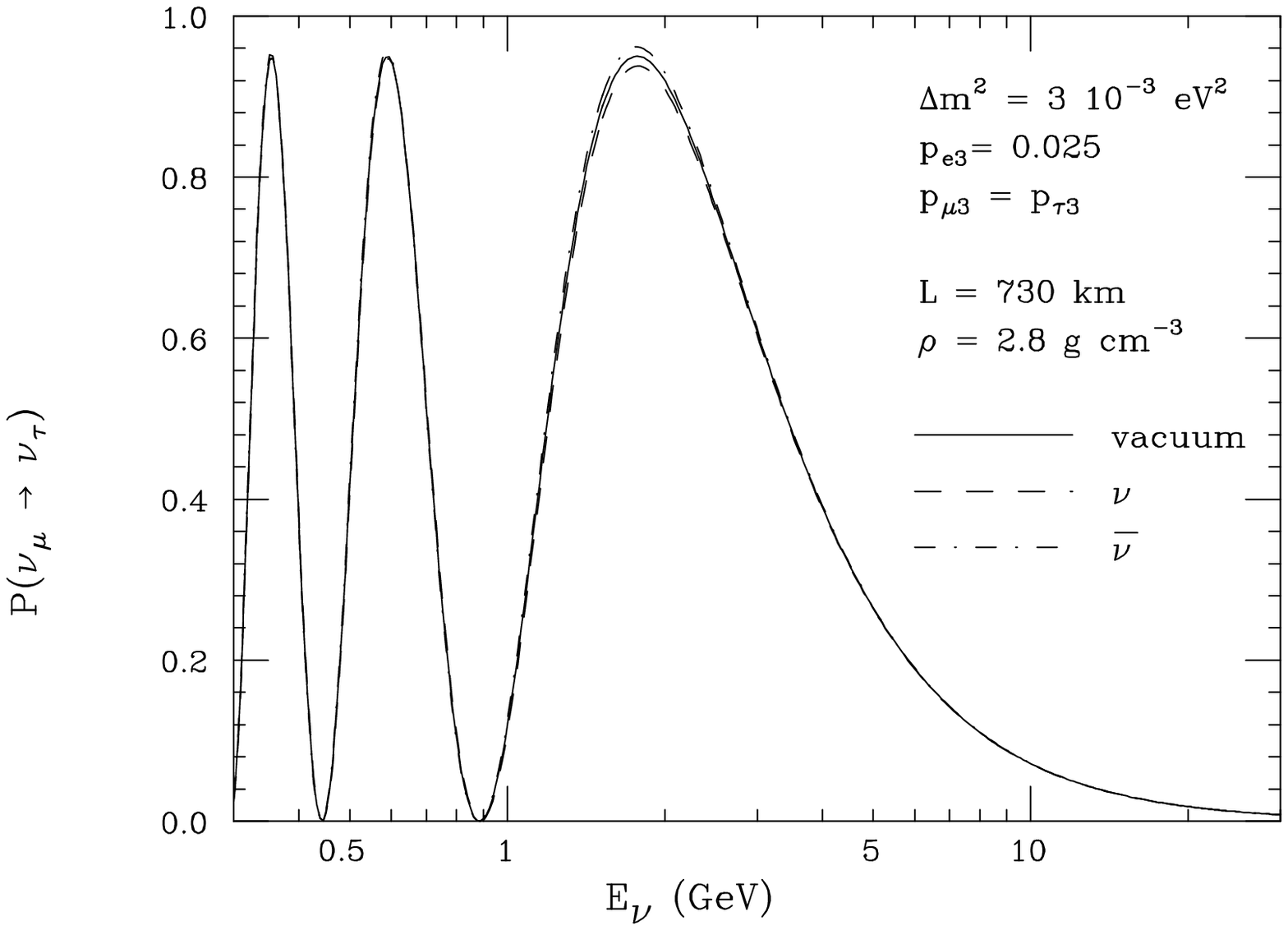,height=12.5cm}}
\caption { Transition probability
$P(\nu_\mu \to \nu_\tau)$ plotted as  a function of
$E_\nu$ for neutrinos  that  have traveled
a distance $L = 730$~km  in 
vacuum (solid  line)  or in matter  
with constant density 
 $\rho = 2.8$~g~cm$^{-3}$
(dashed  line for neutrinos,  dot--dashed  line for antineutrinos).
The neutrino masses  are $m_1 = m_2$,
$m_3^2  = m_1^2 + 3 \cdot 10^{-3}$~eV$^2$.
The  mixing  matrix  in  vacuum is  determined
by $p_{e 3} = \sin^2 \theta=0.025$  and $p_{\mu 3} = p_{\tau 3}$
($\sin^2 \varphi = 0.5$).
\label{fig:tau}
}
\end{figure}

\begin{figure} [t]
\centerline{\psfig{figure=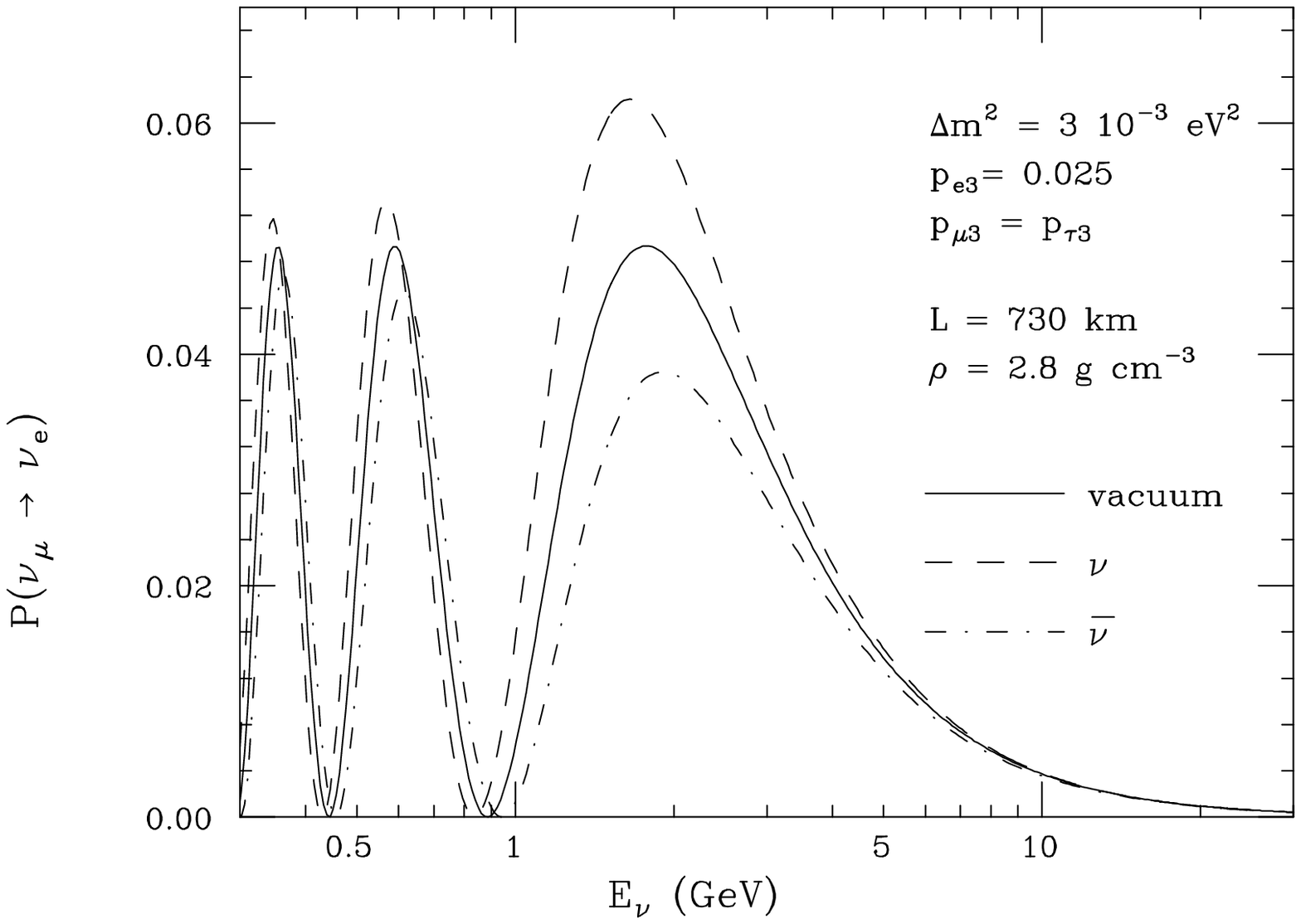,height=12.5cm}}
\caption { Transition probability
$P(\nu_\mu \to \nu_e)$, plotted as  a function of
$E_\nu$ for neutrinos  that  have traveled
a distance $L = 730$~km  in 
vacuum (solid  line)  or in matter  
with constant density 
 $\rho = 2.8$~g~cm$^{-3}$
(dashed  line for neutrinos,  dot--dashed  line for antineutrinos).
The neutrino masses  are $m_1 = m_2$,
$m_3^2  = m_1^2 + 3 \cdot 10^{-3}$~eV$^2$.
The  mixing  matrix  in  vacuum is  determined
by $p_{e 3} = \sin^2 \theta=0.025$  and $p_{\mu 3} = p_{\tau 3}$
($\sin^2 \varphi = 0.5$).
\label{fig:e}
}
\end{figure}

In  fig.~\ref{fig:tau}  and fig.~\ref{fig:e}
as an illustration
we  show the
transition probabilities
$P(\nu_\mu \to \nu_\tau)$
and $P(\nu_\mu \to \nu_e)$
plotted as  a  function of 
$E_\nu$ for neutrinos  that  have traveled
the  distance $L = 730$~km  in 
vacuum (solid  line)  or in matter  
with constant density 
 $\rho = 2.8$~g~cm$^{-3}$
(dashed  line for neutrinos,  dot--dashed  line for antineutrinos).
In these  examples  we have used  $\Delta m^2 = 3 \cdot 10^{-3}$~eV$^2$,
$p_{e 3} = \sin^2 \theta = 0.025$,   and
$p_{\mu 3} = p_{\tau 3} = 0.4875$.
The vacuum transition probabilities  have  a simple
sinusoidal form:
\begin {equation}
 P_{vac} (\nu_\alpha \to \nu_\beta) = 
A_{\alpha\beta}\; \sin^2 \left [ {\pi \over 2} \; {\varepsilon_0 \over E_\nu}
\right ]
\label {eq:simple}
\end{equation}
with maxima  at $E_\nu = \varepsilon_0/n$  ($n$ is a positive
integer)  where the probability has
a  value $A_{\alpha \beta} = 4\, p_{\alpha 3}\, p_{\beta3}$.
The presence of  matter    (for  $p_{e 3} \ne 0$) 
results in   some  deviations of the oscillation
probabilities from the  form (\ref{eq:simple}).
The effect, for  $\Delta m^2 > 0$, is  an enhancement
(suppression)  of the $\nu_\mu \to \nu_e$ 
($\overline {\nu}_\mu \to \overline {\nu}_e$) transition
and correspondingly a suppression  (enhancement)
of the $\nu_\mu \to \nu_\tau$ 
($\overline {\nu}_\mu \to \overline {\nu}_\tau$) transition.
The absolute  size of   $\Delta P = P_{mat} - P_{vac}$ is similar in both
$\nu_\mu \to \nu_e$  and 
$\nu_\mu \to \nu_\tau$  oscillation, but in this second case
the effect is  much more difficult to detect, and less important
to  consider,  because   it represents a small correction
to a  probability of order unity.

Looking at fig.~\ref{fig:tau}  and fig.~\ref{fig:e}
it is possible to observe that the matter  effects  
are  small both  for large  and small  $E_\nu$ and  most important 
in an  intermediate energy.
This  is  true in general and can be easily understood qualitatively.
We can in fact consider three energy regions:
\begin {enumerate}
\item  Large  $E_\nu$   (or  more  precisely
$E_\nu \gg \varepsilon_0$). In this  region   
the oscillation  length is much longer
than  the  pathlength $L$    
($L/\lambda_0(E_\nu) = \varepsilon_0/2 E_\nu$),
and  therefore $P_{mat} \simeq P_{vac}$ for the reasons    described in
section 5.1.

For  the  existing  long baseline  projects,
neutrinos  with energy close to the  MSW resonance 
($E_\nu  \sim \varepsilon_m$)   belong to this  region
since $\varepsilon_m/\varepsilon_0$ is  large,
therefore the presence of  matter does  not result in 
large  visible effects   for   neutrinos  at
the  resonance, even if the   effective masses and  mixing 
are  very  different    from the  vacuum values.

\item  Small $E_\nu$   (or  more  precisely
$E_\nu \ll \varepsilon_m$).
In this region      the   effective  masses
and  mixing of  neutrinos in matter  are close to the vacuum values
($|x| = E_\nu /\varepsilon_m \ll 1$)  and 
again  one has $P_{mat} \simeq P_{vac}$.

\item  Intermediate $E_\nu$.
In this  region the  neutrino  energy
is not   much smaller  that $\varepsilon_m$ {\em  and} 
it is not   much  larger than $\varepsilon_0$.
The first condition 
is needed  to have  significant    modifications of 
on the effective masses and mixing, the second one to 
have  a sufficiently  large  $L/\lambda_0$.
For the  projected  long  baseline  beams 
the  two conditions  can never  be fully  satisfied at the same  time
(because  $\varepsilon_0/\varepsilon_m \le 0.125$ and this is  why the 
matter  effects  never  produce very important effects.
When the neutrino  energy is $E_\nu \sim \varepsilon_0$ 
the conditions  required  to have  significant matter  effects
are  best satisfied, and this is where the matter effects  manifest themselves
most clearly.
\end{enumerate}

\begin{figure} [t]
\centerline{\psfig{figure=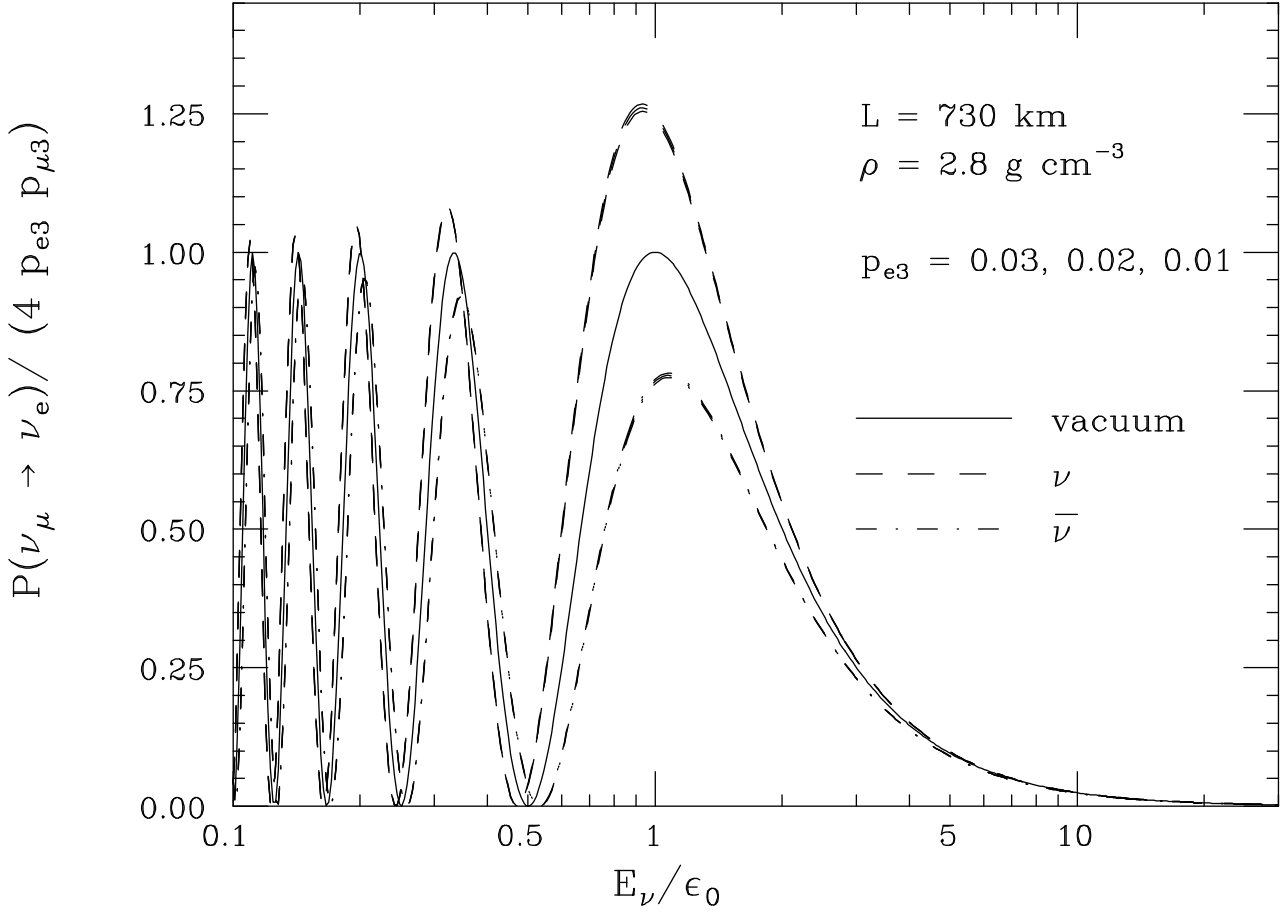,height=12.5cm}}
\caption { Transition probabilities
$P(\nu_\mu \to \nu_e)$, plotted as  a function of
$E_\nu/\varepsilon_0$ for neutrinos  that  have traveled
a distance $L = 730$~km  in  vacuum (solid  line)  or in matter  
with constant density 
 $\rho = 2.8$~g~cm$^{-3}$
(dashed  line for neutrinos,  dot--dashed  line for antineutrinos).
All curves  are valid for   all values of
$p_{\mu 3}$ and  all  positive values of
$\Delta m^2$, for  negative $\Delta m^2$  the  neutrino  and anti--neutrino
curves have to be interchanged.
The   probability    has  beeen calculated for 
$p_{e 3} = \sin^2 \theta = 0.03$, 0.02 and 0.01.
The three  curves for  vacuum   oscillations,
are identical, 
the  curves  calculated taking into account the matter effects  are
very close to each other. 
\label{fig:summ}
}
\end{figure}

Figures~\ref{fig:tau}  and~\ref{fig:e}
have been calculated  for a specific  value of $\Delta m^2$, 
but they describe  oscillations  also  for  an  arbitrary value of 
$\Delta m^2$. In fact 
the oscillation probabilities, for  any  distribution of matter
along the neutrino path, are a function of 
$E_\nu/\Delta m^2$, and the   figures be  considered as 
valid  for  different  values  of 
$\Delta m^2$   simply  rescaling the   neutrino  energy
(for negative  $\Delta m^2$ the probabilities for neutrinos
and  antineutrinos  have to be interchanged).
It  remains to  discuss matter effects for different  neutrino mixings.
This  is  illustrated in  fig.~\ref{fig:summ} that describes the 
transition probability $P(\nu_\mu \to \nu_e)$    for the
Fermilab to Soudan  and CERN  to Gran Sasso projects
for  an  arbitrary value  of $\Delta m^2$  and   for all 
allowed possible mixings.
In the  figure we   plot the probability
$P(\nu_\mu \to \nu_e)$  divided  by  
$4 \,p_{e 3} \, p_{\mu 3}$  (that is the   amplitude of the
oscillation  probability in  vacuum) as  a function of
$E_\nu/\varepsilon_0$ for neutrinos  that  have traveled
a distance $L = 730$~km  in  vacuum (solid  line)  or in matter  
with constant density 
 $\rho = 2.8$~g~cm$^{-3}$
(dashed  line for neutrinos,  dot--dashed  line for antineutrinos).
All curves  are valid for   all values of
$p_{\mu 3}$, and  all  positive values of
$\Delta m^2$ (for  $\Delta m^2 < 0$  the  curve of $\nu$ and $\overline{\nu}$
have to be interchanged).
The   probability    has  beeen calculated for 
three values  of the mixing $p_{e 3} = \sin^2 \theta = 0.03$, 0.02 and 0.01.
The three  curves for  vacuum   oscillations,
are identical.
The  curves  calculated taking into account the matter effects  are
very close to each other, and for  $p_{e 3} \to 0$  they
tend  to an  asymptotic  constant  form.
This    can be  simply  understood observing that
the  $\nu\mu \to \nu_e$  (or $\overline{\nu}_\mu \to \overline{\nu_e}$)
probability can be  written  as:
\begin {equation}
P_{\nu_\mu \to \nu_e} (E_\nu)= { 4 \, p_{\mu 3} \, p_{e 3} \over
   F(x) } ~\sin^2 \left [
 { \Delta m^2 \, \sqrt{F(x)} \; L \over 4 \,E_\nu} \right ],
\end{equation}
where $x = \pm E_\nu/\varepsilon_m$  (the plus (minus)
sign is for $\nu$'s ($\overline{\nu}$'s)),
and   $F(x) = 
\sin^2 2 \theta + (x - cos 2  \theta)^2$.
When   $p_{e 3} = \sin^2 \theta$ is small  (as   indicated by the combined 
analysis  of Chooz and  Super--Kamiokande),
$F(x)$ can be well approximated
as $F(x) \simeq (1 -x)^2$, and 
the   ratio  $P(\nu_\mu \to \nu_e)/p_{e 3}$
becomes  independent from   $p_{e 3}$:
\begin{equation}
{P_{\nu_\mu \to \nu_e} (E_\nu) \over 4\, p_{e 3} \,p_{\mu 3}  } \simeq
  {1 \over (1 - x)^2 } ~ \sin^2 \left [ {\pi \over 2} 
~{\varepsilon_0 \,(1 - x) \over E_\nu} \right ]
\end{equation}
The main effect of matter on $P(\nu_\mu \to \nu_e)$ 
is an enhancement or suppression of 
the probability at the first  maximum $P^*_{\nu(\overline{\nu})}$:
\begin {equation}
 P_{\nu(\overline{\nu})}^*  \simeq  P^*_{vac} 
\left [ 1 \mp 
 {\varepsilon_0 \over \varepsilon_m}  \right ]^{-2} 
\end{equation}
where $P^*_{vac} = 4 \,p_{\mu 3}\, p_{e 3}$.
The minus (plus)  sign  refers to  $\nu$'s ($\overline{\nu}$'s)
for  positive $\Delta m^2$   or viceversa for  negative $\Delta m^2$.
The matter  effects result also  in a displacement of the 
energy $E_{\nu(\overline{\nu})}^*$ of  the first maximum.
Numerically  (assuming   $\Delta m^2 > 0$) for 
the Fermilab  and CERN project  ($L= 730$~km) one has:
\begin {equation}
 P_\nu^* \simeq  P^*_{vac}  \cdot  1.26,  ~~~~~~~~
 P_{\overline{\nu}}^*  \simeq  P^*_{vac} \cdot 0.76
\end{equation}
\begin {equation}
 E_{\nu}^* \simeq \varepsilon_0 \cdot 0.92,
 ~~~~~~~~
 E_{\overline{\nu}}^* \simeq \varepsilon_0 \cdot 1.07
\end{equation}
For  the K2K project ($L= 250$~km)
one  finds:
\begin {equation}
 P_\nu^* \simeq  P^*_{vac}  \cdot  1.09,  
~~~~~~~~
 P_{\overline{\nu}}^*  \simeq  P^*_{vac} \cdot 0.92
\end{equation}
\begin {equation}
 E_{\nu}^* \simeq \varepsilon_0 \cdot 0.98,
 ~~~~~~~~
 E_{\overline{\nu}}^* \simeq \varepsilon_0 \cdot 1.02
\end{equation}

\section {Discussion and conclusions}
The   effect of  matter  on neutrino  oscillations
in  the  projected  long-baseline  neutrino  beams
is  small   but  detectable, especially for the 
longer  pathlength  and  higher  intensity 
Fermilab to Soudan  and  CERN to  Gran Sasso  beams.

In these experiments, assuming $\Delta m^2$ is  positive, 
the effect of matter  will    manifest itself  as an approximately 
25\% enhancement   (suppression)
of  the $\nu_\mu \to \nu_e $ 
($\overline{\nu}_\mu \to \overline{\nu}_e$) probability
in the  crucial region of the first maximum.
The    energy $E_\nu$  where the probability has the first maximum
will   also  be   lower (higher)  than in  vacuum
by  approximately  7\%.
In the shorter  pathlength  K2K  project   the matter effects 
will   result in smaller  effects: a 9\% enhancement of the
probability  and  a 2\%  displacement  of the energy of the first maximum.

It is interesting  to   observe  that  
the   detectable effects   of matter on the oscillations
are   most important not when  $E_\nu$ 
is  close to  the  MSW resonance,
but at  lower  energy when the   
modifications  induced   by  matter  on the   neutrino 
effective masses and  mixing  are smaller, 
but  when   the oscillations  can  develop   because
of  a shorter oscillation  length   
(comparable with $L$) and the  modifications  of the
oscillation parameters  can   produce visible effects.

If $\Delta m^2$,  in  contrast  with the expectations, 
is  negative,  then  the  effects  of matter 
on neutrinos and  anti--neutrinos are  reversed.   
This is  perhaps  the most interesting   effect,
because     it   provides  a  method to measure the  sign of $\Delta m^2$
resolving    the existing  ambiguity.
In order to  do this  one needs to  produce 
(during different    periods of data taking)
beams  of both neutrinos  and anti--neutrinos 
inverting the polarity of the focusing  system    downstream of the
target  region,  and   to detect 
both   $\nu_\mu \to \nu_e$ and 
 $\overline{\nu}_\mu \to \overline{\nu}_e$  transitions,
measuring    the  oscillation  probabilities   for  energies
$E_\nu \sim \varepsilon_0$.  The ratio
$P(\nu_\mu \to \nu_e)/P(\overline{\nu}_\mu \to \overline{\nu}_e)$
can  have only two  possible  values
(approximately 1.64 or 0.61  for the Fermilab and CERN projects,
1.18  or 0.85 for the K2K project)  depending on the sign
(positive  or negative) of $\Delta m^2$.

\vspace {0.8 cm}
\noindent {\bf Acknowledgments}

I would like to thank Maurizio Lusignoli and Giuseppe  Battistoni
for  discussions and  encouragement.

\newpage

\end{document}